\pdfoutput=1

\documentclass[floatfix,aps,prx,reprint,groupedaddress,superscriptaddress,amsmath,amssymb]{revtex4-2}

\usepackage[T1]{fontenc}
\usepackage[utf8]{inputenc}
\usepackage{bm}
\usepackage[hidelinks]{hyperref}
\usepackage{comment}
\usepackage[capitalize,nameinlink]{cleveref}
\usepackage{braket}
\usepackage{siunitx}
\usepackage{graphicx}
\graphicspath{{figs/}}
\begin{document}


\title{Spectral Signatures of Nontrivial Topology in a Superconducting Circuit}



\author{L. Peyruchat}
\author{R. H. Rodriguez}
\affiliation{Quantronics Group, Université Paris Saclay, CEA, CNRS, SPEC, 91191 Gif-sur-Yvette Cedex, France}
\author{J.-L. Smirr}
\affiliation{$\Phi_{0}$, JEIP, USR 3573 CNRS, Collège de France, PSL University, 11, place Marcelin Berthelot, 75231 Paris Cedex 05, France}
\author{R. Leone}
\affiliation{Laboratoire de Physique et Chimie Théoriques, Université de Lorraine, CNRS, F–54506 Vandœuvre lès Nancy Cedex, France}
\author{Ç. Ö. Girit}
\email{caglar.girit@cnrs.fr}
\affiliation{Quantronics Group, Université Paris Saclay, CEA, CNRS, SPEC, 91191 Gif-sur-Yvette Cedex, France}
\affiliation{$\Phi_{0}$, JEIP, USR 3573 CNRS, Collège de France, PSL University, 11, place Marcelin Berthelot, 75231 Paris Cedex 05, France}




\date{\today}

\begin{abstract}
Topology, like symmetry, is a fundamental concept in understanding general properties of physical systems.
In condensed matter, nontrivial topology may manifest itself as singular features in the energy spectrum or the quantization of electrical properties such as conductance and magnetic flux.
Using microwave spectroscopy, we determine that a superconducting circuit with three Josephson tunnel junctions in parallel can possess degeneracies indicative of \emph{intrinsic} nontrivial topology.
We identify three topological invariants, one of which is related to a hidden quantum mechanical supersymmetry.
Measurements show that devices fabricated in different topological regimes fall on a simple phase diagram which should be robust to junction imperfections and geometric inductance.
Josephson tunnel junction circuits, which are readily fabricated with conventional microlithography techniques, allow access to a wide range of topological systems that may have no condensed matter analog.
Notable spectral features of these circuits, such as degeneracies and flat bands, may find use in quantum information, sensing, and metrology.
\end{abstract}


\maketitle



%


\section{\label{sec:intro}Introduction}


A physical system is topologically nontrivial if it has a discrete property that is robust to perturbations~\cite{thouless1998topological}.
For example the magnetic flux through a superconducting ring is quantized.
The number of flux quanta in the ring is a topological invariant that is conserved as long as the ring is not broken.
The quantized transconductance of a semiconductor in the quantum Hall regime is another topological invariant~\cite{vonklitzing40YearsQuantum2020}.
Some topological invariants may be related to singular spectral features such as band crossings~\cite{berryDiabolical1984}.
Many new invariants, describing different kinds of topological order, have been discovered in condensed matter systems~\cite{wang_topological_2017, moessner_moore_2021}.
In electronic systems the topological invariants often correspond to measurable, discrete, universal electromagnetic quantities, such as flux ($h/e$), charge ($e$), or resistance ($h/e^{2}$)~\cite{schererQuantumMetrologyTriangle2012}.
Both topological spectral features and quantized transport parameters find applications in fields such as metrology~\cite{brun-picardPracticalQuantumRealization2016} and quantum information~\cite{gyenisMovingTransmonNoiseProtected2021}.



Although topological phases in materials arise from symmetries of many-particle Hamiltonians~\cite{chiu_classification_2016}, simple single-particle systems with nontrivial topology also exist, such as a spin-1/2 electron in a magnetic field $B$, where the degeneracy at $B=0$ gives rise to a topological charge~\cite{berryDiabolical1984, garg_berry_2010}.
Circuits with Josephson tunnel junctions, due to their versatility and scalability, have been used for the simulation of well-known topological models~\cite{satzinger_realizing_2021}.
However, these circuits, which can often be described by a few bosonic degrees of freedom, also have \emph{inherent} topological properties.
These properties are derived from the intrinsic Josephson junction Hamiltonian, a function of the flux and charge variables~\cite{voolIntroduction2017}, as opposed to a ``simulated'' one, and may have no analog in real materials~\cite{avronquantum1987,avronadiabatic1988,avronInteger1989,pekolaAdiabatic1999,aunolaConnecting2003,mottonenMeasurement2006,mottonenExperimental2008, leoneCooper2008,leoneTopological2008,erdmanFast2019,herrigCooperpair2022,peyruchatTransconductance2021,fatemiWeyl2021,herrigCooperpair2022,peraltagavenskyMultiterminalJosephsonJunctions2023}.

\begin{figure}[t]
      \includegraphics[width=1\columnwidth]{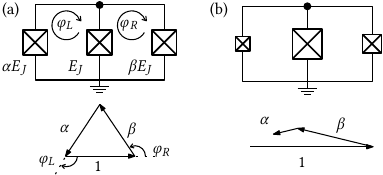}%
      \caption{\label{fig:intro}
        (a) The BiSQUID, a superconducting circuit with three Josephson junctions in parallel, has topologically protected degeneracies when the junctions have comparable Josephson energies $E_J, \alpha E_J, \beta E_J$ such that the triangle condition $\lvert \alpha-\beta \rvert<1<\alpha+\beta$ is fulfilled.
        For large asymmetry (b) the circuit is topologically trivial and has no degeneracies.
        }
\end{figure}

One of the simplest superconducting circuits that has an intrinsic topologically nontrivial phase is the BiSQUID~\cite{griesmar:tel-02345684, fatemiWeyl2021}, composed of three Josephson junctions in parallel~[\cref{fig:intro}].
Using standard circuit quantization, the BiSQUID Hamiltonian can be written in the form,
\begin{equation}
  \label{eq:full_H}
  H = E_C (n - n_g) ^2 - E_J^*(\varphi_L,\varphi_R)\frac{e^{i\theta}}{2}-\bar{E}_J^*(\varphi_L,\varphi_R)\frac{e^{-i\theta}}{2},
\end{equation}
where $n$ and $\theta$ are the canonical Cooper pair number and phase operators  \cite{voolIntroduction2017}.
The charging energy is $E_C = 2e^2/2C$, where $C$ is the total capacitance of the junctions in parallel, and the complex effective Josephson energy is
\begin{equation}
  E_J^*(\varphi_L,\varphi_R) = E_J \left( 1 + \alpha e^{-i\varphi_L} + \beta e^{i\varphi_R} \right).
  \label{eq:Ej_eff}
\end{equation}
The Josephson energy of the middle junction is $E_J$, and $\alpha = E_{JL}/E_J$ and $\beta = E_{JR}/E_J$ are the normalized Josephson energies of the left and right junctions.
The external tunable parameters are the magnetic flux in each loop, $\varphi_{L,R}=2\pi \Phi_{L,R}/\Phi_0$, scaled to the flux quantum $\Phi_0 = h/2e$, and the charge offset $n_g$, scaled by $2e$.
If the effective Josephson energy $E_J^*(\varphi_L,\varphi_R)$ can be tuned to zero by adjusting the fluxes and $n_g$ is integer or half-integer, the BiSQUID spectrum has two distinct degeneracies.
The condition $E_J^*(\varphi_L,\varphi_R)=0$ is equivalent to forming a closed triangle with relative angles $\varphi_L,\varphi_R$ and side lengths $1$, $\alpha$ and $\beta$.
A symmetric BiSQUID ($\alpha=\beta=1$) has an effective Josephson energy that is zero when $\varphi_R=-\varphi_L= \pm 2\pi/3$, corresponding to an equilateral triangle  [\cref{fig:intro}(a)].
When the junction asymmetry is too large, for example $\alpha+\beta<1$, it is no longer possible to close the triangle and the spectrum is nondegenerate [\cref{fig:intro}(b)].
Since the triangle condition can be satisfied for a range of $\alpha, \beta$, the degeneracies are robust to variations in the Josephson energies and hence in the fabrication parameters.
The topological phase diagram is given by the triangle inequality, $|\alpha-\beta|<1<\alpha+\beta$.

In~\cref{sec:exp} we present a microwave spectroscopy experiment to directly determine whether a BiSQUID is topologically nontrivial.
The full theory for the BiSQUID is explained pin~\cref{sec:theory}, where we identify three topological invariants of the Hamiltonian.
The invariants determine whether degeneracies exist (curvature invariant $B$), characterizes the global structure of degeneracies (Witten index $\operatorname{Tr} P$), and associates to each degeneracy a unique winding number $N_2$.
Analysis of the spectra of different BiSQUID devices allows for inferring the topological invariants (\cref{sec:discussion}).

\section{\label{sec:exp}Experiment}

\begin{figure}
    \includegraphics[width=.99\columnwidth]{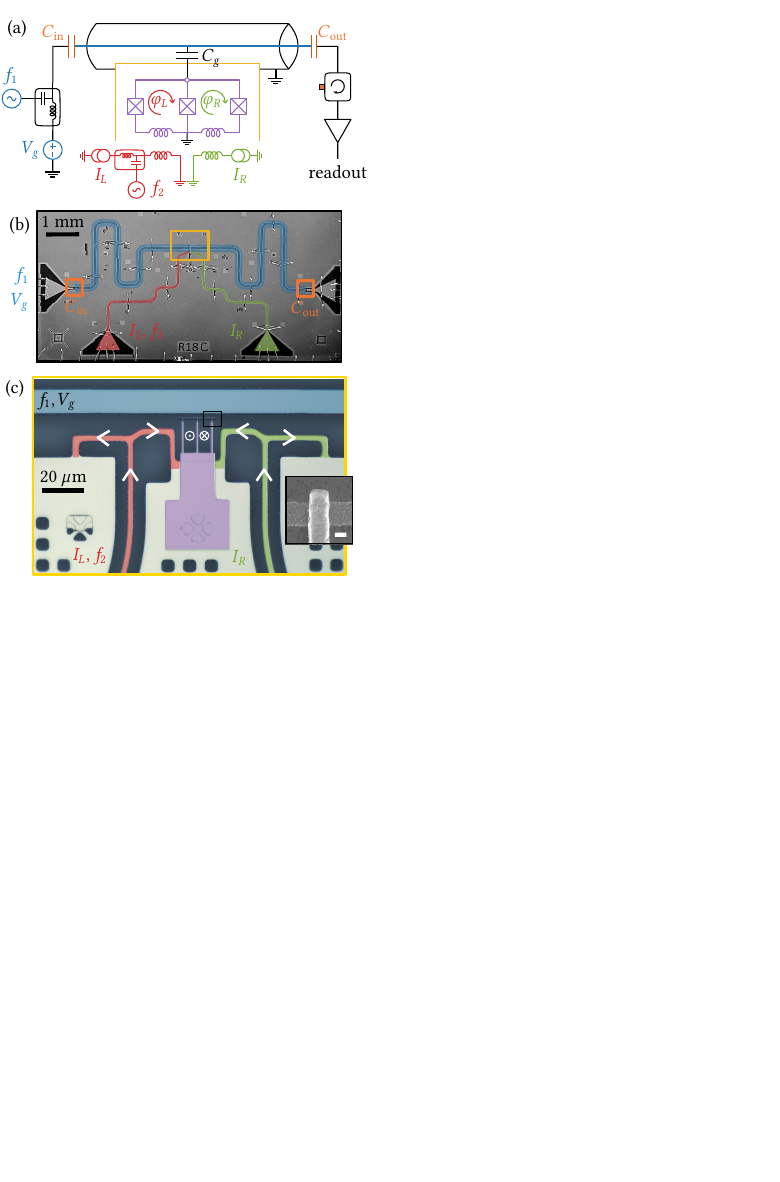}%
    \caption{\label{fig:2}Spectroscopy setup.
    (a) Electrical diagram of a BiSQUID (purple) capacitively coupled to the voltage antinode of a lambda-mode coplanar waveguide resonator (blue). 
    Gate voltage $V_g$ and microwaves at frequency $f_1$ are applied to the resonator using a bias tee (left). 
    Reduced loop fluxes $\varphi_L$ and $\varphi_R$ are set by direct currents $I_L$ and $I_R$ applied to local flux lines (red and green).
    A microwave drive at frequency $f_2$ (left flux line, red) excites transitions of the BiSQUID.
    (b) False-colored optical microscope image of device R18C. 
    The resonator is made of aluminum (gray) on a silicon substrate (black).
    (c) Optical image of BiSQUID, yellow inset in (b).
    Inset: SEM image of rightmost Josephson junction (scale bar $\SI{100}{\nm}$, junction surface area $\SI{0.034}{\micro\meter\squared}$).
    }
\end{figure}

\begin{figure*}
    \includegraphics[width=1\textwidth]{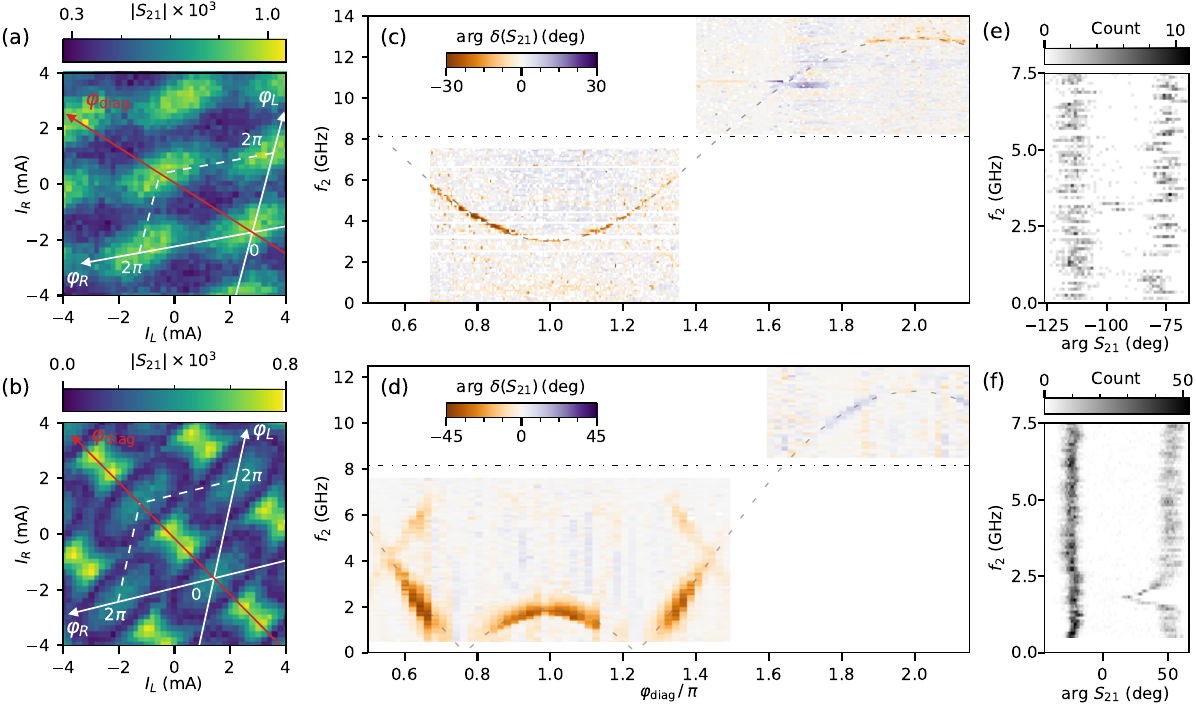}%
    \centering
    \caption{\label{fig:spectro}Spectroscopy of topologically trivial (top row, R17A) and nontrivial (bottom row, R17C) BiSQUIDs at $n_g=1/2$.
      (a),(b) One-tone signal shift of the resonator as a function of currents $I_L,I_R$ in the two flux lines.
      Superposed are reduced flux axes $\varphi_L$ and $\varphi_R$ (solid white lines), diagonal $\varphi_\mathrm{diag}=\varphi_L=\varphi_R$ (red lines), and unit cell (dashed lines). 
    (c),(d) Two-tone phase shift of the resonator along $\varphi_\mathrm{diag}$.
    Overlaid on the data are the bare frequency of each resonator (dash-dotted line) and the transition frequency $f_{01}$ obtained from fitting to the BiSQUID Hamiltonian [\cref{eq:full_H}] (dashed lines).
    (e),(f) Histograms of $\textrm{arg}\, S_{21}(f_2)$ at $\varphi_{\textrm{diag}}=\pi$ representative of data used to obtain spectra in panels (c) and (d).
    The distribution is bimodal due to quasiparticle poisoning.
    }
\end{figure*}

Spectroscopy is a direct probe of energy transitions and is well suited for  identifying topologically nontrivial systems via spectral features.
Circuit quantum electrodynamics~\cite{blaisCircuitQuantumElectrodynamics2021} techniques are routinely used for superconducting qubit characterization and manipulation.
We use a microwave cavity resonator to perform the spectroscopy of several BiSQUIDs in both trivial and topological regimes (\cref{fig:2}).
A BiSQUID (purple) is capacitively coupled to the voltage anti-node of a coplanar waveguide lambda-mode resonator (blue), with $f_r$ approximately \SIrange{7}{9}{\GHz} and a loaded quality factor $Q_L > 10^3$.
We first deposit $\SI{100}{\nm}$ aluminum on a silicon substrate and then fabricate the resonator and control lines using photolithography and wet etching.
The Josephson junctions of the BiSQUID are made of aluminum and are patterned using e-beam lithography and a double-angle evaporation technique.
The surface area of each junction is approximately $200\times\SI{200}{\nm \squared}$ such that $E_J\approx E_C$ [inset, \cref{fig:2}(c)].
All experiments are conducted in a dilution refrigerator at $\SI{10}{\milli\K}$.
The transmission of the resonator at frequency $f_1$ is probed using a vector network analyzer (VNA), and transitions between energy levels of the BiSQUID are detected via changes in $f_1$ resulting from driving the BiSQUID with a second tone of frequency $f_2$ via a local flux line.
Further details of the setup are provided in [\cref{app:expsetup}].

The charge offset $n_g$ is controlled by applying a gate voltage $V_g=2en_g/C_g$ to the central conductor of the resonator.
Even with careful filtering of the dc gate voltage line, the value of $n_g$ slowly drifts a few percent per hour and randomly jumps on a timescale of minutes.
To compensate, we repeatedly perform quick calibrations approximately every minute.
We sweep $V_g$ over a few periods while probing the resonator to find the value of $V_g$ corresponding to the desired $n_g$.
This process allows us to correct for slow drift and detect charge jumps, in which case data are discarded and remeasured.

Quasiparticle poisoning is a large source of error for many superconducting quantum devices \cite{glazmanBogoliubov2021}.
High-energy photons or ionizing radiation \cite{vepsalainenImpact2020,mcewenResolving2021,cardaniReducing2021,gordonEnvironmentalRadiationImpact2022,connollyCoexistenceNonequilibriumDensity2024} may create nonequilibrium quasiparticles which can then tunnel to the superconducting island of the BiSQUID, shifting $n_g$ by 0.5 (1e).
Despite broadband noise filtering and superconducting gap engineering, we measure constant jumps between quasiparticle parities with a typical switching time on the order of $\SI{300}{\ms}$ and equal population of each parity (\cref{app:QP}).
Post-measurement processing allows sorting of the data for each parity state.

Direct currents $I_L$ and $I_R$ are applied to two local flux lines (see red and green in \cref{fig:2}) to control loop fluxes $\varphi_L$ and $\varphi_R$.
The second microwave tone used to excite transitions of the BiSQUID is applied via one of these flux lines.
To find the conversion between applied currents $\left(I_L,I_R\right)$ and reduced fluxes $\left(\varphi_L, \varphi_R\right)$, we determine the unit cell for the periodic resonator shift [\cref{fig:spectro}(a) and 3(b)] and compensate accordingly.

We fabricate several BiSQUIDs with different Josephson energy asymmetries $\alpha$ and $\beta$ to explore the topological phase diagram.
BiSQUID R17A s fabricated with the middle junction about 3 times larger than the outer ones ($\alpha\approx \beta \approx 1/3$), such that the spectrum should be gapped.
Two other BiSQUIDs, R17C and R18C, are fabricated with junctions of similar area in order to have degeneracies.
For simplicity, only the middle junction area is varied, keeping $\alpha\approx\beta$ such that degeneracies should be located on the high-symmetry axis $\varphi_L=\varphi_R \equiv \varphi_\mathrm{diag}$.

A first signature of the topological phase of the BiSQUID is found in the flux maps [\cref{fig:spectro}(a) and 3(b)].
The asymmetric BiSQUID R17A ($\alpha = \beta= 0.34$) shows only a single minimum (dark blue) in one unit cell [\cref{fig:spectro}(a)].
On the contrary, for the more symmetric device R17C ($\alpha = \beta = 0.67$), we observe two maxima (yellow).
The extrema in both maps correspond to local minima of the BiSQUID transition energy.
For a topologically trivial BiSQUID, such as R17A (a), the gap should be smallest at the high-symmetry point $\varphi_\mathrm{diag}=\pi$.
For a BiSQUID in the topologically nontrivial regime, there are always two degeneracies, and thus two extrema.

We then perform two-tone spectroscopy by measuring the transmission $S_{21}$ of the resonator in the presence of a second tone of frequency $f_2$ at different values of $\varphi_\mathrm{diag}$.
The charge offset is set to $n_g=0.5$, one of the two possible values where degeneracies are expected, and postprocessed to account for quasiparticle poisoning.
The two-tone phase shift is shown in \cref{fig:spectro}(c) and 3(d), where the flux-dependent single-tone phase offsets have been removed.
We observe one main spectroscopy line corresponding to the transition $f_{01}$ from the ground to the first excited state.
At $\varphi_\mathrm{diag}=2\pi$, the transition has a maximum of \SIrange{12}{13}{\GHz} for both BiSQUIDs.
However, near $\varphi_\mathrm{diag}=\pi$, the transition strongly differs for the two circuits.
For the BiSQUID in the trivial regime, $f_{01}$ has a single minimum of $\SI{3}{\GHz}$ near $\varphi_\mathrm{diag}= \pi$ [\cref{fig:spectro}(c)].
The BiSQUID in the topologically nontrivial regime shows two minima, at $\varphi_\mathrm{diag}\approx 0.75\pi$ and $\varphi_\mathrm{diag}\approx 1.25\pi$, and the transition has negative curvature and a local maximum at $\varphi_\mathrm{diag}= \pi$.
The $f_{01}$ spectral line disappears below $\SI{1}{\GHz}$ because the charge-coupled resonator shift goes to zero at the degeneracy point~\cite{parkAdiabaticDispersiveReadout2020}.

The raw data leading to the maps in \cref{fig:spectro}(c) and 3(d) are shown for the respective devices at a flux bias $\varphi_{\textrm{diag}}=\pi$ in \cref{fig:spectro}(e) and 3(f).
Each horizontal cut is a histogram of the phase of the transmitted signal  $\textrm{arg}\,S_{21}$ at the resonator frequency obtained from multiple measurements while driving at the second tone of frequency $f_2$ ($y$ axis).
The color scale corresponds to the number of counts.
Because of quasiparticle poisoning, the distribution is bimodal, with the second, spurious, mode corresponding to a charge offset $n_g \pm 0.5 = 0$.
To properly resolve both quasiparticle parity states, individual measurements are averaged over a time window of $\SI{1}{\ms}$, much smaller than the quasiparticle tunneling time of $\SI{100}{\ms}$ or more. 
A threshold is applied to distinguish between the two states, and only data corresponding to $n_g=0.5$ are shown in \cref{fig:spectro}(c) and 3(d).
The desired mode, $n_g=0.5$, shows clear shifts when $f_2$ is resonant with a BiSQUID transition, whereas the spurious mode is relatively constant.
This is because the excitation gap near $\varphi_{\textrm{diag}}=\pi$ for $n_g \equiv 0$ is above approximately \SI{8}{\GHz}.
The outlined procedure allows us to measure the BiSQUID transitions even in the presence of quasiparticle poisoning. 

\section{\label{sec:theory}Theory}

\begin{figure*}
  \includegraphics[width=1\textwidth]{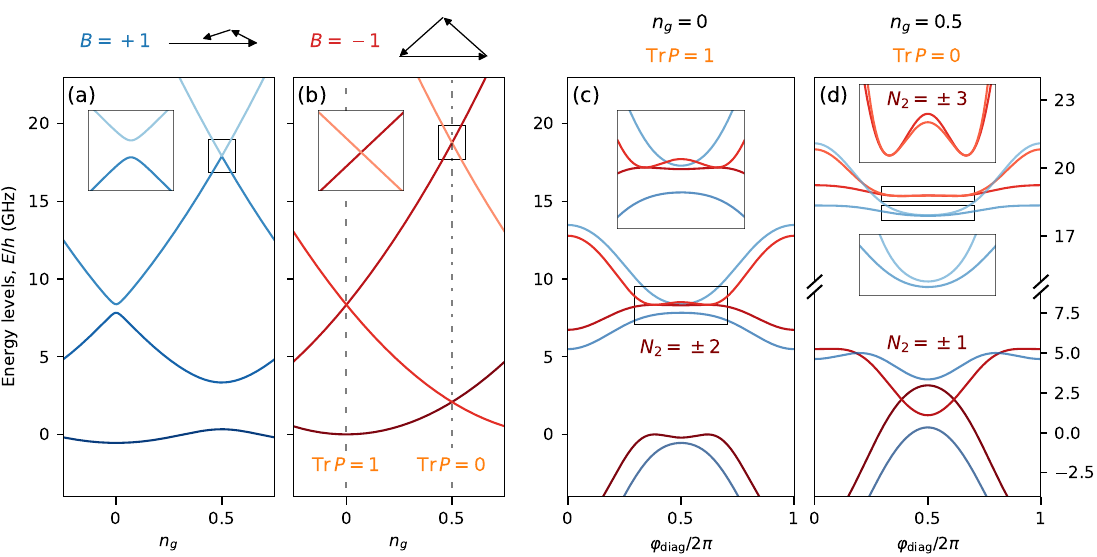}%
  \caption{\label{fig:topohier}
    (a) When a BiSQUID does not satisfy the triangle condition the spectrum is gapped.
    The plot parameters are $\varphi_L=\varphi_R=\pi$, which minimize the gaps.
    The associate topological invariant is $B = +1$.
    (b) When the triangle condition is satisfied the effective Josephson coupling can be tuned to zero and the spectrum has degeneracies ($B = -1$).
    The existence of degenerate excited states at $n_g = 0$ and $n_g = 1/2$ (dashed vertical lines) is indicative of a hidden supersymmetry, characterized by the topological invariant $\operatorname{Tr} P$, the Witten index.
    (c) Plot of the energy spectrum along $\varphi_L=\varphi_R=\varphi_{\textrm{diag}}$ for the two supersymmetric configurations, showing two band crossings in the Brillouin zone for $B = -1$ (red), each associated with a third topological invariant, the winding number $N_2$.
    Note that $B = +1$ is gapped (blue).
    The wavefunction continuity is indicated by the line color.
  }
\end{figure*}

To understand the spectroscopy data of~\cref{fig:spectro}, we describe the topological invariants of a BiSQUID and relate them to the measured spectral features.
The invariant $B$ encodes the satisfiability of the triangle condition~[\cref{fig:intro}], which is equivalent to the existence of zeros of the effective Josephson potential $E_J^*(\varphi_L,\varphi_R)$ [\cref{eq:Ej_eff}].
When the Josephson energies of the three junctions of the BiSQUID are such that the triangle condition cannot be satisfied ($B = +1$) [\cref{fig:topohier}(a)], the spectrum as a function of the charge offset $n_{g}$ is gapped for all fluxes $\varphi_{L,R}$.

On the other hand if the triangle condition can be fulfilled ($B = -1$)  [\cref{fig:topohier}(b)], $E_J^*(\varphi_L,\varphi_R) = 0$ at two distinct values of $\varphi_{L,R}$ where the spectrum reduces to that of the charging Hamiltonian $H_{C} = E_C(n-n_{g})^{2}$ \cite{bouchiat_quantum_1998,nakamura_coherent_1999}.
The eigenstates are $\ket{n}$, with $n$ integer, corresponding to the number of excess Cooper pairs on the ungrounded electrode.
The Hamiltonian $H_{C}$ can also describe the flux states of a superconducting ring, a quantum rotor, or the angular momentum states of a charged particle confined to a ring~\cite{robnik_false_1986, aharonov_aharonov-bohm_1994, meijer_one-dimensional_2002, liu_quantum_2003, vugalter_charged_2004}.

The energy-level diagram of the charging Hamiltonian [\cref{fig:topohier}(b)] has a peculiar degeneracy structure.
At $n_{g} = 0$, all states except the zero-energy ground state, are doubly degenerate, whereas at $n_{g}=\pm1/2$, all states are degenerate.
This spectral signature is reminiscent of supersymmetric quantum mechanics (SUSY QM), a toy model for SUSY quantum field theories in which a bosonic Hamiltonian is augmented with an additional fermionic ``superpartner'' degree of freedom~\cite{nicolaiSupersymmetrySpinSystems1976, witten_dynamical_1981, lahiri_supersymmetry_1990, cooper_supersymmetry_1995}.
Whereas the charging Hamiltonian $H_{C}$ has no explicit fermionic component, it has a hidden supersymmetry that gives rise to a fermionic degree of freedom (\cref{app:susy}).
Hidden supersymmetry has been considered in several different quantum bosonic systems: the free particle~\cite{rauSupersymmetryQuantumMechanics2004}, the bound state Aharonov-Bohm problem~\cite{correaHiddenSupersymmetryQuantum2007, jakubskyOriginHiddenSupersymmetry2010}, Jaynes-Cummings model~\cite{andreevSupersymmetryJaynesCummingsModel1989}, and superconducting circuits~\cite{ulrichSimulationSupersymmetricQuantum2015b}.

The generic supersymmetric Hamiltonian can either have a doubly degenerate spectrum with a zero-energy ground state singlet, called unbroken SUSY ($n_g = 0$), or a doublet nonzero-energy ground state, called broken SUSY ($n_g = \pm 1/2$).
The topological invariant associated with supersymmetry is the Witten index $\operatorname{Tr} P$ [\cref{fig:topohier}(b)], which sums the parities $P$ of all states in the different supersymmetric configurations, with $\operatorname{Tr} P = 1$ for unbroken SUSY ($n_g=0$) and $\operatorname{Tr} P = 0$ for broken SUSY ($n_g=1/2$).

Thus, for the BiSQUID, when the triangle inequality is satisfied ($B = -1$), one can have either broken or unbroken SUSY depending on the value of $n_g$, resulting in a spectrum with a tower of degeneracies [\cref{fig:topohier}(b), dashed vertical lines].
Although only a few pairs of degenerate excited states are shown, all other excited states will also be doubly degenerate at the same diabolical points~\cite{berryDiabolical1984} in the $(\varphi_L,\varphi_R)$ Brillouin zone [\cref{fig:levels-topo}].

In the analogy to a topological material, the reduced magnetic fluxes $\varphi_{L,R}$ that are odd under time reversal play the role of $k_x,k_y$ for a 2D crystal.
The offset charge $n_g$, induced by an applied electric field, is even under time-reversal and therefore considered a control parameter~\cite{fatemiWeyl2021}.
In the condensed matter context, the BiSQUID at $n_g=0$ or $1/2$ is a 2D topological semimetal with time-reversal and inversion symmetry.
The conical two-band crossings are referred to as either 2D Weyl points~\cite{fengTwodimensionalTopologicalSemimetals2021d} or Dirac points of spinless graphene (class AI with inversion)~\cite{manesExistenceTopologicalStability2007,chiu_classification_2016}.

In the topologically nontrivial regime one can also assign a topological charge to each degeneracy.
This local topological charge, the winding number $N_2$~\cite{volovikQuantumPhaseTransitions2007}, is calculated by integrating the Berry phase around the degeneracy in the $\varphi_{L,R}$ plane at $n_g = 0$ or $1/2$, where the system is time-reversal invariant.
If we consider the Brillouin zone as three dimensional, including $n_g$ as a parameter in addition to $\varphi_{L,R}$, the same topological charge can be calculated by integrating the Berry curvature on a surface enclosing the degeneracy and is referred to as the Chern number~\cite{berryDiabolical1984,garg_berry_2010, griesmar:tel-02345684, peraltagavenskyMultiterminalJosephsonJunctions2023}.
The two states of a degenerate doublet have opposite winding numbers, and at $n_g = 0$ [\cref{fig:topohier}(c)], $N_2$ is always even and starts at $N_2=\pm2$ for the first excited-state degeneracy.
For $n_g=1/2$ [\cref{fig:topohier}(d)], $N_2$ is odd and starts at $N_2=\pm1$ for the ground-state degeneracy.

For a BiSQUID with $B = -1$, the degeneracies are topologically protected in the sense that, since the triangle condition is a set of inequalities, one can change the Josephson energies of the junctions and charging energy without immediately opening a gap.
The zeros of $E_J^*(\phi_L, \phi_R)$ will shift with small changes in $\alpha,\beta$; hence, the locations of the degeneracies will move in the $(\phi_L, \phi_R)$ plane, but the degeneracy will not be lifted~\cite{griesmar:tel-02345684,fatemiWeyl2021}.

Because of the Von Neumann-Wigner theorem~\cite{vonneumanUber1929}, barring accidental degeneracies, at least three parameters are necessary to obtain band crossings in a generic Hamiltonian.
This requirement explains why the BiSQUID, with one charge offset and two flux parameters, is the minimal parallel Josephson circuit with protected degeneracies.
A gate-tunable two-junction superconducting quantum interference device (SQUID), or Cooper-pair transistor, only has a single flux parameter in addition to the charge offset.
In a perfectly symmetric SQUID, with $E_{JL} = E_{JR}$, the effective Josephson energy will be zero when the flux is biased at half a flux quantum ($\Phi = \Phi_0/2$), leaving the charging Hamiltonian $H_C$.
However, this result requires two junctions with exactly identical critical currents which is impossible to fabricate.
Any asymmetry in the Josephson energies of a SQUID will lift the degeneracies at $n_g=0, \pm 1/2, \Phi = \Phi_0/2$.

\section{\label{sec:discussion}Discussion}

\begin{figure}[h]
    \includegraphics[width=.99\columnwidth]{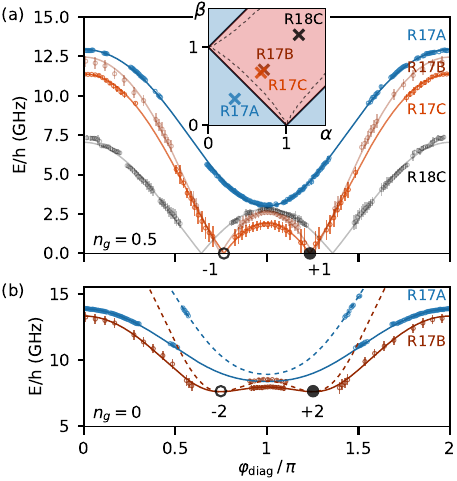}
    \caption{\label{fig:result}
      BiSQUID transitions $f_{01}$ (circles along solid lines) and $f_{02}$ (circles along dashed lines) as a function of $\varphi_{\textrm{diag}}$ extracted from two-tone spectroscopy measurements at $n_g=1/2$ (a) and $n_{g}=0$ (b) for different devices.
      Error bars indicate the measured FWHM of transitions, and lines are best fits to the data using the BiSQUID Hamiltonian.
      The inset shows the phase diagram with placement of devices in topologically nontrivial (red) and trivial (blue) regions, separated by the critical line (solid), which is shifted (dashed) when taking into account geometric inductance, $\beta_{L}=0.1$ (\cref{app:betal}).
      Labels at degeneracies indicate the winding number $N_2$.
    }
\end{figure}

The transition frequencies extracted from raw data, as in [\cref{fig:spectro}], are plotted as a function of $\varphi_\mathrm{diag}$ for four BiSQUID devices, R17A, R17B, R17C, and R18C (\cref{fig:result}).
The error bars correspond to the full width at half maximum (FWHM) of the Gaussian fits.

We fit the transitions measured as a function of $\varphi_{\textrm{diag}}$ for $n_g=1/2$ and $n_g=0$ to the eigenenergies of the BiSQUID Hamiltonian~\cref{eq:full_H} using three free parameters: $E_C$, $E_J$, and the outer junction symmetry parameter $\alpha=\beta$.
These fits are plotted as solid lines ($f_{01}$) and dashed lines ($f_{02}$) in~\cref{fig:result}, and they show good agreement with the data (circles).
For comparison with raw data, the fit of $f_{01}$ for R17A and R17C is overlaid in \cref{fig:spectro}(c) and 3(d).

The circuit parameters obtained from fits, along with properties of the resonator, are given in \cref{tab:all_param} and are consistent with geometric estimates based on scanning electron microscopy.
As a control, we also performed spectroscopy of a SQUID (device \emph{R16}), which was designed to be symmetric, and fit to a model in which $\beta = 0$ (\cref{app:sqdata}).
Fitting yields a gap for the BiSQUID in the topologically trivial regime and the control SQUID.
In the case of the SQUID, the gap is due to a residual junction asymmetry of $2\%$ resulting from fabrication variations.

\begin{table}[h]
  \centering
  \begin{tabular}{lccccc}
    \hline
                           &  R17A         & R17B           & R17C          & R18C            & \emph{R16}  \\ \hline
    $f_r$ (GHz)            & 8.096         & $9.001$        & $8.171$       & $7.245$         & $9.178$       \\
    $Q_L$                  & 37 520        & 4 200          & 65 460        & 17 520          & 5020          \\
    $Q_\mathrm{ext}$       & 185 000       & 75 000         & 185 000       & 50 000           & 7400          \\
    $E_{J}/h$ (GHz)       & 9.37          & $6.24$         & $5.53$        & 2.15            &  7.56             \\
    $ \alpha$              & 0.34          & 0.71           & 0.67          & 1.16            & 1.02        \\
    $ \beta$               & 0.34          & 0.71           & 0.67          & 1.16            & 0        \\
    $E_C/h$ (GHz)          & 7.93          & $7.61$         & $8.34$        & 16              & $9.05$         \\
    $\omega_p/2\pi$ (GHz)  & 15.78         & $15.17$         & $14.72$      & 15              & $16.61$       \\
    $E_{01,\mathrm{min}}/h$ (GHz) & 3.0    & 0              & 0              & 0                & 0.2       \\ \hline
  \end{tabular}
  \caption{\label{tab:all_param}
    Summary of parameters of BiSQUID devices and control device SQUID \emph{R16} [\cref{app:sqdata}].
  }
\end{table}

As in~\cref{fig:spectro}, the most indicative qualitative feature of a topologically nontrivial regime at $n_g=1/2$ is the existence of negative curvature in $f_{01}$ near $\varphi_{\mathrm{diag}}=\pi$, as observed in samples R17B, R17C, and R18C~[\cref{fig:result}(a)].
This feature is in contrast to sample R17A (blue), which has a spectrum with positive curvature at $\varphi_{\mathrm{diag}}=\pi$ reminiscent of an asymmetric SQUID.
We will show that the band curvature, measured at $\varphi_{\mathrm{diag}}=\pi$, is equivalent to the topological invariant $B$ and directly indicates whether a BiSQUID is topologically nontrivial.

The phase diagram for the BiSQUID is given by the triangle inequality $|\alpha-\beta|<1<\alpha+\beta$ [\cref{fig:result}(a), inset], and it determines whether there are solutions to $E_J^*(\varphi_L,\varphi_R) = 0$ \cite{weisbrichTensor2021}.
The two degeneracy points will move in the Brillouin zone as the junction asymmetries $\alpha$ and $\beta$ are varied, converging at high-symmetry points where $\varphi_L$ and $\varphi_R$ are either zero or $\pi$ as the parameters approach the critical lines $|\alpha\pm\beta|=1$.
On the critical lines, the greatest Josephson energy is equal to the sum of the other two, and the diabolical points merge to produce an energy dispersion described by a power law.

The topologically nontrivial device, sample R17C, with $\alpha=\beta\approx0.67$, is closest to the critical line $\beta=1-\alpha$.
In this case, the degeneracy triangle is more oblique, with angles $\varphi_{L,R}$ (\cref{fig:intro}) closer to $\pi$.
Hence, the region of negative curvature is smaller, more closely confined to a region around $\varphi_{\textrm{diag}}=\pi$.

The unbroken SUSY regime, $n_g=0$, is shown in \cref{fig:result}(b).
Data of the topologically nontrivial device R17B (red) show negative curvature at $\varphi_{\textrm{diag}}\approx \pi$ for both transitions $f_{01}$ and $f_{02}$, which is in agreement with the model, plotted as a red solid line ($f_{01}$) and a dashed line  ($f_{02}$).
The negative curvature in the topologically nontrivial state is in contrast to the expectation for device R17A, where the fit lines for $f_{01}$ and $f_{02}$ do not intersect and both transitions would have positive curvature near $\pi$.
For this sample, there are no data at $\varphi_{\textrm{diag}}\approx \pi$ because the resonator frequency is coincident with the transition frequency.
In the unbroken SUSY state ($n_g=0$), the lack of data where $f_{01}$ is degenerate with $f_{02}$ at $\varphi_{\textrm{diag}}\approx 0.75\pi$ (indicated by $\pm2$ winding number) is due to the vanishing of the resonator frequency shift, similarly to what occurs for the ground-state degeneracy in broken SUSY ($n_g=1/2$)~\cite{parkAdiabaticDispersiveReadout2020}.

As with topological insulators, negative band curvature at a high-symmetry point can be indicative of an inverted gap.
Indeed, the phase diagram is described by a topological invariant that relates the curvature of the ground-state band at high-symmetry points to the validity of the triangle inequality.
At the four inequivalent high-symmetry points $(\varphi_L,\varphi_R)=(0,0),(0,\pi),(\pi,0),(\pi,\pi)$ of the 2D Brillouin zone, there is time-reversal symmetry, and $E_J^*(\varphi_L,\varphi_R)$ is real.
As shown in~\cref{app:curvinv}, the triangle inequality can be fulfilled only if one of the three points $[(0,\pi),(\pi,0),(\pi,\pi)]$ has negative $E_J^*$.
The $\mathbb{Z}_2$ ``curvature'' invariant $B$ is defined as the product of the signs of $E_J^*$ evaluated at these points.
From the Hamiltonian~\cref{eq:full_H}, the sign of the expectation value of $\cos\theta$ in the ground state, $\operatorname{sgn} \braket{\cos\theta}_0$, must be equal to the sign of $E_J^*$ at each of the symmetry points in order to minimize the energy.
Therefore, $B$ can also be calculated from the product of the signs of the ground-state expectation values $\braket{\cos\theta}_0$ over the symmetry points.

Using the Hellmann-Feynman theorem, it can be shown that for $\alpha = \beta$, the curvature of an energy band in the $\varphi_{\textrm{diag}}$ direction at $\varphi_{\textrm{diag}}=\varphi_L=\varphi_R=\pi$ is proportional to the expectation value of $\braket{\cos\theta}$ (\cref{app:curvinv}).
The sign of the curvature measured for the transition energy $f_{01}$ at $\varphi_{\textrm{diag}}=\pi, n_g=1/2$ is then related to $\operatorname{sgn} \braket{\cos\theta}_0$.
Furthermore, for the experimentally realized devices where $\alpha,\beta < 1$ (R17A, R17B, and R17C), it suffices to determine the sign of $\braket{\cos\theta}_0$ at the unique point $\varphi_{\textrm{diag}}=\pi$ to determine $B$.
Hence, the sign of the curvature at $\varphi_{\textrm{diag}}=\pi$ of the measured spectra, \cref{fig:result}(a), is a direct indication of whether the device is topologically nontrivial.

In topological insulators with inversion and time-reversal symmetry, there is a similar $\mathbb{Z}_2$ invariant calculated from the product of parity eigenvalues over the high-symmetry points~\cite{fuTopologicalInsulatorsInversion2007}.
For the BiSQUID, $B$ is not a band invariant since it is only defined for the ground-state, but it encodes information about the junction asymmetries and therefore the triangle inequality.

In the broken SUSY configuration, \cref{fig:result}(a), the topological charge of the ground-state degeneracy is either $+1$ or $-1$, as indicated by black circles.
In unbroken SUSY, \cref{fig:result}(b), the topological charges of the degeneracies between states $\ket{1}$ and $\ket{2}$ are $\pm 2$, as indicated by black circles.
Higher-order degeneracies have correspondingly higher charges but are odd for $n_g=1/2$ and even for $n_g=0$~\cite{herrigCooperpair2022}.
Along the critical line of the phase diagram separating trivial and nontrivial states, the degeneracies merge and the integral of the Berry curvature around the band intersection is zero, equal to the sum of the topological charges.

Higher experimental sensitivity would allow the measurement of transitions closer to the degeneracy points at $n_g=0$ and $n_g=1/2$.
However, spectroscopic techniques can only give an upper bound to a possible energy gap.
Thermal effects, sample disorder, and instrument limitations set bounds on the emission linewidth and detector resolution, ultimately determining the smallest transition that can be measured.
Controversy over the existence of a gap occurred previously in angular-resolved photoemission studies of epitaxial graphene~\cite{zhouSubstrateinducedBandgapOpening2007,zhouOriginEnergyBandgap2008,rotenbergOriginEnergyBandgap2008}.

Despite the fact that the data are consistent with a Hamiltonian [\cref{eq:full_H}] that has a nontrivial phase, one cannot exclude that unknown, perturbative terms open a gap in the region of the spectrum that we cannot resolve experimentally.
Although many possible mechanisms could introduce such terms, including capacitive nonlinearities~\cite{herrigQuasiperiodicCircuitQuantum2023a}, we only discuss the role of Josephson harmonics and the loop inductance.

Higher-order Josephson effects are important in short, transparent, superconducting weak links, and they add harmonic terms of the type $\cos k\theta$ for $k > 1$, corresponding to the simultaneous tunneling of multiple Cooper pairs.
These processes, which are related to multiple Andreev reflections, are always weaker than the single Cooper pair tunneling process.
Although they are strongly suppressed in typical tunnel junctions, they can occur when there are defects~\cite{willschObservationJosephsonHarmonics2024}.
Increased junction transparency can be incorporated in the effective Josephson energy [\cref{eq:Ej_eff}] by adding exponential terms with arguments $ik\varphi_{L,R}$ with $k > 1$.
The triangle condition for degeneracy will be modified but can still be satisfied.
Considering the symmetry point $\varphi_L=\varphi_R=\pi$, the even and odd harmonic terms in the effective Josephson energy will sum with opposite signs, in effect making it harder to ``close'' the triangle when even harmonics are present.
However one can compensate by increasing the total Josephson energy of the outer junctions.
The size of the nontrivial region of the phase diagram will therefore shrink as the junction transparency increases and higher-order Josephson terms become more important, but it will not disappear completely.

The superconducting loops of the BiSQUID will have a small inductance, both geometric and kinetic, which can be accounted for by adding terms of the form $E_L\gamma^2= E_J/\beta_L\gamma^2$, where $\beta_L$ is the ratio of the loop inductance to the Josephson inductance and $\gamma$ the phase drop across the inductance.
In the perturbative limit $\beta_L \ll 1$, inductances effectively reduce the Josephson energy of the junctions by a factor $\cos \rho \gamma \approx 1-(\rho \gamma)^2/2$, where $\rho$ is a coefficient of order unity depending on the geometry (\cref{app:betal}).
Replacing $\gamma^2$ by its expectation value for an $l$-photon oscillator state, the renormalization factor is approximately $1-\rho^2\frac{2l+1}{4}\sqrt{E_C/E_L}$.
Therefore, when accounting for inductance, the low-lying degeneracies of a BiSQUID will not be lifted if the BiSQUID is deep in the topologically nontrivial regime, far from the critical lines.
However, for highly excited BiSQUID states, the oscillator degrees of freedom may be in a large number state $m \gg 1$, and it will be harder to satisfy the triangle inequality.

Solving numerically for the energy levels in the presence of arbitrary loop inductance (\cref{app:betal}), we find that although the lowest-lying degeneracies persist, both for integer and half-integer $n_g$, higher-order degeneracies may split according to their topological charge or even merge and open a gap~\cite{pintér2022birth}.
The deformation of the phase diagram for the ground-state degeneracy at $n_g=1/2$, due to artificially large series inductances $\beta_L=0.1$ on each branch of a BiSQUID with $E_J=E_C=\sqrt{E_{JL}E_{CL}}=\sqrt{E_{JR}E_{CR}}$, is shown in the inset of \cref{fig:result}(a).
For small $\beta_L$, the prediction of the perturbative model for the shift in the phase diagram of the position of the critical line near the point $\alpha=\beta=1/2$ agrees with numerical results~(\cref{app:betal}).
Similarly to the effect of higher-order Josephson terms, the inductive perturbation reduces the size of the topologically nontrivial region.
However, whereas the phase diagram accounting for junction transparency is identical for all excited states, leading to a supersymmetric spectrum [\cref{fig:topohier}(b)] in the topologically nontrivial state, the inductive terms will gap out states at high enough energy.
Accounting for inductance, supersymmetry is effectively destroyed in the ultraviolet limit.
The role of additional bosonic degrees of freedom in lifting supersymmetry at the superconducting critical point of fermionic Dirac-Weyl systems has been considered previously~\cite{zhaoAbsenceEmergentSupersymmetry2019}.

\section{Conclusion}

The BiSQUID is a model topological system that has at least three different topological invariants: the curvature invariant $B$, the Witten index $\operatorname{Tr} P$, and the winding number $N_2$.
The curvature invariant, which is a function of the relative Josephson junction energies, determines whether a given BiSQUID sample can have degeneracies.
For topologically nontrivial devices $B = -1$, the structure of degeneracies is characterized by the Witten index, and the winding number classifies individual degeneracies.
We have measured the curvature invariant of several different BiSQUIDs via microwave spectroscopy, demonstrating that it is possible to identify a topologically nontrivial state without having to directly probe degeneracies.
The spectra are fit to a model with a supersymmetric degeneracy structure from which we infer the winding numbers of the degeneracies.

In future experiments, both the resonator-BiSQUID coupling and BiSQUID excitation should be modified to optimize the dispersive shifts near level crossings.
It is also desirable to excite high-order level crossings, demonstrating that they occur at the same bias point, establishing the defining characteristic of SUSY.
This process requires pumping transitions at higher frequencies, pumping with multiple drive tones, using alternative techniques such as amplitude spectroscopy~\cite{bernsAmplitudeSpectroscopySolidstate2008}, and optimizing the Josephson junction plasma frequency and resonator frequency.
Single-tone measurements with a low-frequency resonator should directly give the ground-state band curvature and could be a powerful probe of circuit topological properties.
With three-tone spectroscopy, one may also be able to probe transitions from excited states, such as $f_{12}$.

By measuring the transition energies along multiple flux axes and determining the Gaussian curvature at $\varphi_{\textrm{diag}}=\pi$, it may be possible to obtain the topological invariant $B$ for the general case $\alpha \ne \beta$.
The numerically calculated energy bands indicate a saddle point or negative Gaussian curvature for topologically nontrivial devices.
Other pump-probe spectroscopic techniques~\cite{gritsevDynamicalQuantumHall2012,ozawaProbingLocalizationQuantum2019,kleesMicrowaveSpectroscopyReveals2020} should be able to directly determine the winding number of degeneracies.

We have shown how the topological properties are sensitive to realistic perturbations.
Whereas the higher-order Josephson effect does not affect supersymmetry, inductance can open gaps where highly excited states would otherwise be degenerate.
Increasing the loop inductance $\beta_L$ with granular aluminum~\cite{Maleeva2018} or other high kinetic inductance materials may allow observing the lifting of supersymmetry.
Neglecting the role of perturbations can lead to the misinterpretation of experimental results and tenuous claims of topological nontriviality~\cite{dasSarmaSmokingGun,frolov2023smoking}.
Further theoretical work should elucidate how the supersymmetric spectrum at $n_g=0$ and $n_g=1/2$ is modified as $\beta_L$ is increased and predict the splitting and merging of high-order topological charges.

Experiments on devices with an additional junction and one more loop, the TriSQUID, would effectively allow \emph{in situ} tuning of the junction asymmetry $\alpha$~\cite{leo_thesis,fatemiWeyl2021}.
One could then traverse the phase diagram, switching between topologically trivial and nontrivial regimes, with an additional local flux line.
Crossing the critical line would allow one to observe the merging of Weyl points and the opening of a gap~\cite{monjouMergingDiabolicalPoints2013b}.
With more sophisticated circuits, one could perform Weyl-point teleportation~\cite{frankWeylpointTeleportation2024} or probe higher-order topological invariants~\cite{weisbrichTensorMonopolesSuperconducting2021b}.
Another approach to dynamically probe the phase diagram is to use gate-tunable multiterminal Josephson devices~\cite{pankratovaMultiterminalJosephsonEffect2020,grazianoSelectiveControlConductance2022} instead of tunnel junctions.
The Josephson energy of semiconductor-superconductor weak links can be directly tuned with electric fields.
Such weak links have already been incorporated into a BiSQUID geometry~\cite{coraiolaPhaseengineeringAndreevBand2023} with local gates, which would allow for adjusting $\alpha$ and $\beta$ and exploring the phase diagram.

It would also be interesting to investigate, both theoretically and experimentally, the existence of supersymmetry in circuits with many junctions~\cite{fodaSupersymmetricPhaseTransition1988a} or circuits with nontunnel junctions, which harbor Andreev states~\cite{bretheauExcitingAndreevPairs2013}.
Single~\cite{zazunovAndreevLevelQubit2003,janvierCoherentManipulationAndreev2015b,tosiSpinOrbitSplittingAndreev2019b,haysCoherentManipulationAndreev2021} or multiterminal~\cite{vanheckSingleFermionManipulation2014c,riwarMultiterminal2016,peraltagavenskyMultiterminalJosephsonJunctions2023} circuits with Andreev states may have degeneracies described by supersymmetry~\cite{galaktionovSupersymmetricHamiltonianSolutions2020}.

With respect to topological order in condensed matter systems, the material analog of a BiSQUID is 2D spinless graphene.
It remains to establish a general mapping between the topological classes of materials~\cite{chenSymmetryProtectedTopological2013b,chiu_classification_2016,zengSymmetryProtectedTopologicalPhases2019} and Josephson-junction-based circuits.
Inversion symmetry plays an important role in the BiSQUID, and it may be possible to establish the link between the ``weak'' BiSQUID invariants, $\operatorname{sgn}\braket{\cos\theta}$ at high-symmetry points, the ``strong'' invariant $B$, and the corresponding topological invariants of inversion symmetric fermionic topological insulators~\cite{fuTopologicalInsulatorsInversion2007}.
Sublattice symmetries (\cref{app:susy}) can also lead to degeneracies, as in Weyl semimetals and Josephson junction circuits with higher-order terms~\cite{egusquizaRoleAnomalousSymmetry2022}.
Although one can define a Brillouin zone for the BiSQUID, it does not have a boundary in real space, and therefore it has no edge states.
On the other hand, it is possible to use Josephson junctions to produce high-dimensional topological systems without any real-world analog~\cite{fatemiWeyl2021}.
A full classification of the topological properties of arbitrary Josephson junction circuits is lacking~\cite{avronadiabatic1988,leo_thesis}.

The existence of nontrivial topology, manifest here as degeneracies in the spectrum of a superconducting circuit, can be exploited in applications.
Several proposals to realize transconductance quantization with Josephson tunnel junctions incorporate BiSQUIDs~\cite{peyruchatTransconductance2021,weisbrichFractionalTransconductanceNonadiabatic2023a}.
The dual circuit of the BiSQUID, the Cooper pair pump, which also has nontrivial topology~\cite{leoneCooper2008, leoneTopological2008}, has already been developed for current metrology~\cite{pekolaSingleelectronCurrentSources2013a, erdmanFastAccurateCooper2019}.
In both classes of circuits, degeneracies are essential to obtain voltages, currents, or resistances quantized in terms of the fundamental constants $h$ and $e$~\cite{schererQuantumMetrologyTriangle2012,poirierAmpereElectricalUnits2019}.
Similar spectroscopic techniques as employed here for BiSQUIDs could help evaluate the suitability of the proposed circuits for electrical metrology.
Circuits similar to the BiSQUID can also serve in quantum information applications.
Topological properties can help mitigate decoherence or decay~\cite{ioffeTopologically2002,doucotTopological2003,gladchenkoSuperconducting2009,usmanovProtected2010,gyenisMovingTransmonNoiseProtected2021,brookesProtectionQuantumInformation2022}, and enable platforms for geometric and holonomic computation~\cite{falciDetection2000,solinasGroundstate2010,zhangGeometricHolonomicQuantum2023}.

\appendix
\label{appendix}
\section{\label{app:susy}Supersymmetry of Charging Hamiltonian}

\begin{figure}
    \includegraphics[width=\columnwidth]{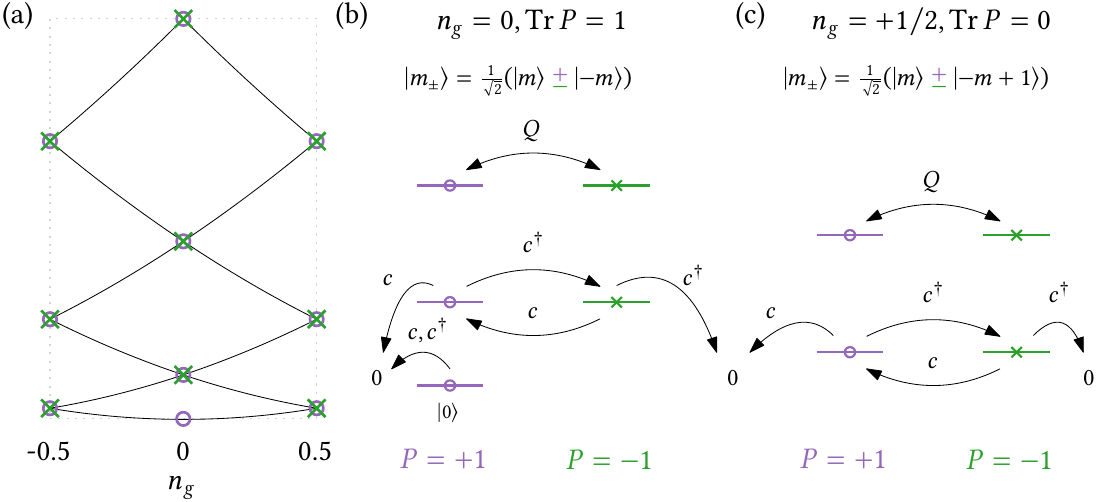}%
    \caption{\label{fig:susy_diagram}
        (a) Hidden quantum mechanical supersymmetry in the spectrum of the charging Hamiltonian.
        (b) At charge offset $n_g=0$ (Witten index $\operatorname{Tr} P = 1$), supersymmetry is \emph{unbroken}: the ground state at zero energy is not degenerate, whereas all excited states are doubly degenerate.
        (c) At $n_g = \pm 1/2$ ($\operatorname{Tr} P = 0$), supersymmetry is \emph{broken}: the ground state energy is greater than zero and all states are doubly degenerate.
        ``Bosonic'' (even parity, $P = +1$) and ``fermionic'' (odd parity, $P = -1$) superpartners are indicated by purple circles and green crosses.
      The supercharge operator $Q$ exchanges eigenstates of even parity $\ket{m_+}$ (bosonic) and odd parity $\ket{m_-}$ (fermionic) for $m_{\pm}>0$.
      The fermion number operator $c^{\dag}c$ annihilates bosonic states while leaving fermionic states invariant.
      Eigenstates are written in charge basis $\ket{m}$ satisfying $Q\ket{m}=\sqrt{E_C}(m-n_g)\ket{m}$.
    }
\end{figure}

Here we consider the hidden supersymmetry, both broken and unbroken, inherent in any superconducting circuit where the Hamiltonian at degeneracy is the charging Hamiltonian $H_C$~\cite{rauSupersymmetryQuantumMechanics2004}.
\cref{fig:susy_diagram} is a diagram of the quantum mechanical supersymmetry properties of the charging Hamiltonian $H_C$ at $n_g=0$ and $n_g=+1/2$.

The symmetry that partitions eigenstates of $H_C$ into ``bosons,'' with even parity, and ``fermions,'' with odd parity, is inversion about the point $n = n_{g}$.
In the charge basis, the parity operator is $P(n_{g}) = \sum_{n}\ket{-n+2n_{g}}\bra{n}$, defined only for $n_{g}=0,\pm 1/2$ when restricted to the Hilbert space.
Since $[P, H_C] = 0$ and $P^2 = 1$, energy eigenstates can also be chosen to be parity eigenstates with eigenvalues $\pm 1$.
Combining $P$, referred to as a grading operator in supersymmetry, with the ``supercharge'' operator, $Q(n_{g}) =\sqrt{E_{C}}(n-n_{g})$, we have the following algebra,
\begin{align}
  \label{eq:susy}
  P^{2} = 1,\qquad \{P,Q\} = 0,\qquad Q^{2} = H_{C},
\end{align}
which is an equivalent definition of SUSY QM~\cite{combescureAreSupersymmetricQuantum2004}.
These relations imply that all nonvanishing eigenvalues of $H_C$ are associated with degenerate eigenvectors and give rise to the energy-level structure of~\cref{fig:topohier}(b).
The singlet ground state at $n_{g} = 0$ lacks a ``fermionic'' partner because there is no odd-parity eigenstate of zero momentum.

The simultaneous parity and energy eigenstates of $H_C$ are given by $\ket{m_{\pm}} = \frac{1}{\sqrt{2}}(\ket{m}\pm\ket{-m+2n_g})$ where $m > 0$, except for the unique ground state at $n_g=0$ which is given by $\ket{0}$.
Note that charge eigenstates $\ket{\pm m}$ are distinct from energy eigenstates $\ket{m_{\pm}}$.
For unbroken SUSY, $n_g=0$, the real wave functions are sines and cosines with wave vector $m$, whereas for broken SUSY, $n_g=1/2$, the complex wave functions have a magnitude given by sines and cosines with wave vector $m/2$.

For $n_g=0$ and $n_g=+1/2$ we therefore have $P\ket{m\pm}=\pm\ket{m\pm}$ and $H_C\ket{m\pm}=E_m\ket{m\pm}$ with $E_m=E_C(m-n_g)^2$.
The positive sign corresponds to even parity, or bosonic states, and the negative sign to odd parity, or fermionic states.

The supersymmetry operator $Q = \sqrt{E_{C}}(n-n_{g})$ swaps parity, or exchanges the bosonic and fermionic states, up to a multiplicative factor, $Q\ket{m_{\pm}} = \sqrt{E_C}(m-n_g)\ket{m_{\mp}}$.
By construction, we obtain the fundamental spectral property of supersymmetry: for an eigenstate $\ket{m_{\pm}}$ with nonzero energy, $Q\ket{m_{\pm}}$ is a \emph{distinct} degenerate eigenstate.
The uniqueness of the two states ultimately arises from the anticommutator $\{P,Q\}=0$~(\cref{eq:susy}).
When a symmetry operator $S$ commutes with the Hamiltonian, unless the operator is antiunitary, the eigenstates $\ket{\psi}$ and $S\ket{\psi}$ may be equivalent up to a phase factor.
However, since both $P$ and $Q$ commute with the Hamiltonian \emph{and} the anticommutator $\{P,Q\}$ is zero, one can show that for any nonzero energy eigenstate $\ket{\psi}$, either $P\ket{\psi}$ or $Q\ket{\psi}$ is also an eigenstate and differs from $\ket{\psi}$.

For the bosonic charging Hamiltonian, the supersymmetry is hidden, in the sense that there is no explicit fermionic degree of freedom.
However we can identify a fermionic operator $c = \frac{1+ P}{2}Q$, which swaps parity and then projects states to the even, or bosonic sector.
Since $\{P,Q\} = 0$, the adjoint $c^\dag = \frac{1-P}{2}Q$ also swaps parity but projects to the odd sector.
The operators $c,c^{\dag}$ are fermionic in the sense that they satisfy $c^2=c^{\dag 2}=0$ and $\{c,c^{\dag}\}/E_m=1$ for all eigenstates with non-zero energy.
When applied to a fermionic state $\ket{m_-}$, the fermionic number operator $c^{\dagger}c$ is the identity (up to a constant), but $c^{\dagger}c$ annihilates a bosonic state $\ket{m_+}$, hence the particle-naming convention for the different parities.

Using the fermion number operator or parity operator, one can partition the Hilbert space in two, and rewrite the Hamiltonian in terms of Pauli matrices~\cite{combescureAreSupersymmetricQuantum2004}.
This process yields a supersymmetric Hamiltonian in the same way that, in conventional SUSY QM, a Hamiltonian with a generic potential but no explicit fermionic component is supplemented by its superpartner~\cite{cooper_supersymmetry_1995,rauSupersymmetryQuantumMechanics2004}.

The Witten index $\operatorname{Tr} P$ sums the parities of all states, counting the difference between the number of bosonic and fermionic states: $1$ for unbroken SUSY ($n_{g} = 0$) and $0$ for broken SUSY ($n_{g} = \pm 1/2$).
Note that $\operatorname{Tr} P$ is ill-defined for noninteger $2n_g$ as the $2\pi$ periodicity of the wave functions restricts $n$ to integer values.


A normalized analogue to the superexchange operator can be defined as $q = Q/\sqrt{H_C} = (n-n_g)/|n-n_g| = \operatorname{sgn}(n-n_g)$, which is unitary if we take the convention $\operatorname{sgn}(0)=+1$ and not $0$. 
Like $P$, $q$ is a grading operator that squares to the identity and partitions states by the sign of the charge relative to $n_g$.
Note that with unitary $q$, since $q\ket{0} = \ket{0}$, the fundamental SUSY anticommutator $\{q,P\}$ is zero for all eigenstates except the ground state at $n_g=0$.
For $n_g=1/2$, since the eigenstates $\ket{m_{\pm}}$ mix even and odd charge states, the superexchange operator can also be written as $q = e^{i\pi n}$, which, in the position basis, is a half-lattice translation.

Degeneracies in the spectrum can also be understood as arising from Kramers theorem.
The antiunitary symmetry $PqK$, where $K$ is the time reversal operator (complex conjugation in the charge basis), squares to $-1$ for all eigenstates except the ground state at $n_g=0$.
Therefore, all nonzero energy eigenstates are Kramers pairs.
This statement is equivalent to the SUSY degeneracy theorem.
At $n_g=1/2$, $Pq$ can be considered a glide symmetry, since $P$ is a reflection and $q = e^{i\pi n}$ is a half-lattice translation.
Such sublattice symmetry also arises in Dirac-Weyl systems~\cite{armitageWeyl2018}.

For the five-junction unbroken supersymmetric circuit proposed by Ulrich \emph{et al.}~\cite{ulrichSimulationSupersymmetricQuantum2015b}, the grading operator is the same as for $n_g=0$, but the supercharge has an additional ``nontrivial'' term proportional to $\sin\theta$.
At a special value of the proportionality constant, all states except the ground state are doubly degenerate.
Unfortunately, the supersymmetry is not robust to imperfections in the fabrication process, and the circuit is more difficult to realize experimentally.

Supersymmetry also plays a role in models such as SYK~\cite{behrendsSupersymmetryStandardSachdevYeKitaev2020a}; it has been debated to exist in certain topological superconductors~\cite{groverEmergentSpaceTimeSupersymmetry2014,zhaoAbsenceEmergentSupersymmetry2019} and has been measured in artificial systems such as ion trap simulators~\cite{caiObservationSupersymmetryIts2022}.

\section{\label{app:curvinv}Curvature Invariant}

\begin{figure}
    \includegraphics[]{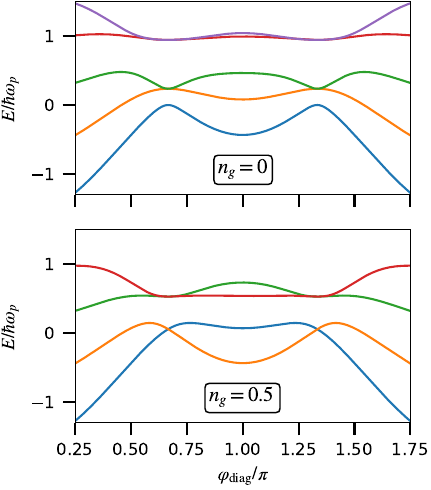}%
    \caption{\label{fig:levels-topo}
      Plot of the first few energy levels of a fully symmetric, topologically nontrivial BiSQUID with $E_C=E_J/3$ and $\alpha=\beta=1$.
      When $\varphi_{\textrm{diag}}=\pm2\pi/3\approx \pm 0.67\pi$, the triangle condition is satisfied and the effective Josephson energy $E_J^{*}$ is zero.
      In the unbroken SUSY case ($n_g=0$, top) all states except the ground state at zero energy are doubly degenerate.
      For broken SUSY ($n_g=1/2$, bottom), all states are doubly degenerate.
    Line colors determined by wave function continuity \cite{akhmerovConnectingDots2017}.
  }
\end{figure}

We show that the curvature invariant $B$, the product of $\operatorname{sgn} E_J*$ over the high-symmetry points, encodes the triangle condition in~\cref{fig:intro} by enumerating its value for all possible configurations of $\alpha, \beta$ (\cref{tab:Binv}).
At the four high-symmetry points $(\varphi_L,\varphi_R)=(p\pi,q\pi)$ where $p,q = 0$ or $1$ (first column), the effective normalized Josephson energy $E_J^{*}(\varphi_L,\varphi_R)$ \cref{eq:Ej_eff} can be written $1+(-1)^p\alpha+(-1)^q\beta$ (second column).
Assuming that $\alpha \ge \beta$ without loss of generality, there are three possible orderings of the Josephson energies relative to $E_J$: $1\le \beta\le \alpha$, $\beta<1\le \alpha$, and $\beta\le \alpha < 1$ (top row).
Except for the last row, the cells with $+,-,\pm,\mp$ give the sign of $E_J^{*}$ for the corresponding high-symmetry point and Josephson energy ordering.
For the cells with two possibilities, the sign of $E_J^{*}$ needs to be resolved by using the triangle inequality: whether the sum of the smallest of $1,\alpha,\beta$ is greater than the largest one.
For the orderings in columns three and four, the triangle inequality is $1+\beta \ge \alpha$, and for the last column, it is $\alpha+\beta \ge 1$.
For the double entries indicated by $\pm$ or $\mp$, the top sign corresponds to the Josephson energy configuration that satisfies the triangle inequality.

The last row gives the product of the signs along a column, which is equivalent to the topological curvature invariant $B$.
In all columns, the product is negative if and only if the upper sign in $\pm$ or $\mp$ is taken or, equivalently, when the triangle inequality is satisfied.
Therefore, independent of the Josephson energy ordering, $B=-1$ implies that the triangle inequality can be satisfied.

\begin{table}[t]
  \centering
\begin{tabular}{c|c|c|c|c}
  \centering
  & & \multicolumn{3}{c}{$\operatorname{sgn}E_J^{*}$} \\
  $\varphi_L,\varphi_R$ & $E_J^{*}$ & $1 \le \beta \le \alpha$ & $\beta<1 \le \alpha$ & $\beta \le \alpha <1$ \\
  \hline\hline
  $(0,0)$ & $1+\alpha+\beta$ & $+$ & $+$ & $+$ \\
  $(\pi,0)$ & $1-\alpha+\beta$ & $\pm$ & $\pm$ & $+$ \\
  $(0,\pi)$ & $1+\alpha-\beta$ & $+$ & $+$ & $+$ \\
  $(\pi,\pi)$ & $1-\alpha-\beta$ & $-$ & $-$ & $\mp$ \\
  \hline\hline
  \multicolumn{2}{c|}{$\Pi \operatorname{sgn}E_J^{*}$} & $\mp$ & $\mp$ & $\mp$
\end{tabular}
  \caption{\label{tab:Binv}
    Calculation of curvature invariant $B$.
  }
\end{table}

To relate $B$ to the curvature $E_j''$ of the $j$th energy band along the direction $\varphi_{\textrm{diag}}$ for the mirror symmetric configuration $\alpha=\beta$, we first expand the flux dependence of the effective Josephson energy $E_J^{*}(\phi_L,\phi_R)$ in the Hamiltonian~\cref{eq:full_H},
\begin{equation}
    \label{eq:full_H_X}
    \begin{aligned}
      H = &E_C (n - n_g) ^2 -E_J\cos \theta \\
      &-E_J\alpha \cos(\theta - \varphi_L) - E_J\beta \cos(\theta + \varphi_R).
    \end{aligned}
\end{equation}
The potential energy term is rewritten as a function of flux variables $\sigma = (\varphi_R+\varphi_L)/2$ and $\delta = (\varphi_R-\varphi_L)/2$, $-E_J\left (1+2\alpha\cos \sigma \cos (\theta+\delta)\right )$.
With this transformation, the diagonal $\varphi_{\textrm{diag}}=\varphi_R=\varphi_L$ in the Brillouin zone~\cref{fig:spectro}(a) is equivalent to $\sigma = \varphi_{\textrm{diag}}, \delta = 0$.
Therefore the prime notation $'$ refers to derivatives with respect to $\sigma$.

Taking the derivative of the Hellmann-Feynman theorem $E_j' = \bra{\psi_j}H'\ket{\psi_j}$, where $\ket{\psi_j}$ is the $j$th energy eigenstate, we obtain an expression for the curvature,
$$E_j'' = \bra{\psi_j}H''\ket{\psi_j} + \bra{\psi_j'}H'\ket{\psi_j} + \bra{\psi_j}H'\ket{\psi_j'}.$$
At the high-symmetry point $\varphi_{\textrm{diag}}=\pi$, or $\sigma = \pi, \delta = 0$, the last two terms are zero as long as $\alpha = \beta$.
Hence, $E_j''/E_J = -2\alpha \braket{\cos\theta}_j$, and the curvature is proportional to $\braket{\cos\theta}_j$, the expectation value of $\cos\theta$ in state $\ket{\psi_j}$.

All measured devices except R18C have Josephson energies satisfying the inequality $\alpha\approx\beta < 1$.
According to the last column of~\cref{tab:Binv}, for $\alpha,\beta<1$ the invariant $B$ will be given by the sign of the effective Josephson energy at $\varphi_{\mathrm{diag}}=\pi$ (last row) or, equivalently, the sign of $\braket{\cos\theta}$ in the ground state, $\braket{\cos\theta}_0$.
From the numerical surface plots of the first two energy manifolds [insets of~\cref{fig:spectro}(c) and (d)], one sees that $E_0''$ is expected to be positive for the topologically nontrivial configuration ($B = -1$) and negative in the trivial configuration ($B = 1$).
This finding concurs with $\operatorname{sgn} E_0'' = -\operatorname{sgn} \braket{\cos\theta}_0 = -B$ as inferred from the formula $E_j''/E_J = -2\alpha \braket{\cos\theta}_j$.

Since~\cref{fig:result} shows the transition energy $f_{01}$, the curvature of the data at $\varphi_{\textrm{diag}}=\pi$ should be compared to $E_1''-E_0'' = -2\alpha E_J(\braket{\cos\theta}_1-\braket{\cos\theta}_{0})$.
At $\varphi_{\mathrm{diag}}=\pi$ and for $n_g=0$ or $n_g=1/2$, the parity operator $P$ commutes with the Hamiltonian, and energy eigenstates are also parity eigenstates.
Wave functions constructed from the $n_g=1/2$ parity basis, $\ket{m\pm} = 1/\sqrt{2}\left( \ket{-m}\pm\ket{m+1} \right)$ (\cref{app:susy}), always have a node at either $\theta = \pi$ (positive parity) or $\theta = 0$ (negative parity).
Therefore, the expectation value $\braket{\cos\theta}_1$ of the first excited state, which has opposite parity compared to the ground state, cannot be larger in absolute value than $\braket{\cos\theta}_{0}$.
This requirement implies that $\operatorname{sgn}(E_1''-E_0'') = -\operatorname{sgn}(\braket{\cos\theta}_1-\braket{\cos\theta}_{0}) = \operatorname{sgn}\braket{\cos\theta}_0 = B$.

As stated in the main text, the sign of the curvature measured at the high-symmetry point $\varphi_{\mathrm{diag}}=\pi, n_g=1/2$ is a direct indication of $B$ for all samples except R18C, which has $\alpha\approx\beta>1$.
For R18C, looking at~[\cref{tab:Binv}], the curvature is always negative at $\varphi_{\mathrm{diag}}=\pi$, and a second measurement of the curvature at $\varphi_L = \pi, \varphi_R=0$ would be needed to resolve the triangle inequality.
However, in practice, since we know that $\alpha \approx \beta \approx 1.16$ for R18C [\cref{tab:all_param}], we expect that the triangle inequality $1+\beta\ge \alpha$ will be satisfied and that a measurement of the curvature at $\varphi_L = \pi, \varphi_R=0$ would have positive sign.
The blue curve in~\cref{fig:result}(a) ($n_g=1/2$) for the gapped device R17A has positive curvature ($B = 1$) at $\varphi_{\mathrm{diag}}=\pi$ whereas the two gapped devices (red) have negative curvature ($B = -1$).

\section{\label{app:betal}Role of Inductance}


The BiSQUID Hamiltonian~\cref{eq:full_H_X} can be modified to incorporate inductances on each branch,
\begin{align*}
  H_l = H_1& (n_1,\theta_1) +H_2(n_2,\theta_2) + E_{C\Sigma}(n-n_g)^2+ \\
  -E_J  [\, &\cos (\theta+\gamma_1\theta_1+ \gamma_2\theta_2) + \\
  \alpha &\cos(\theta + \gamma_3\theta_1+ \gamma_4\theta_2 - \varphi_L) + \\
  \beta &\cos(\theta + \gamma_5\theta_1+\gamma_6\theta_2+\varphi_R) ],
\end{align*}
where $H_i(n_i,\theta_i)$ are harmonic oscillator Hamiltonians describing the two normal modes of the linear part of the circuit composed of branch capacitors and inductors.
The coordinates of the normal modes are the conjugate quantum variables $n_i,\theta_i$.
Each mode has frequency $\omega_i$ and dimensionless impedance $z_i$, which is the characteristic impedance scaled by a resistance $\hbar/(2e)^2$.
The constants $\gamma_i < 1$ are participation ratios~\cite{minevEnergyparticipationQuantizationJosephson2021}, and they describe the coupling of the harmonic modes to the Josephson junction phases.
The effective Josephson degrees of freedom $n,\theta$ are associated with a common mode with charging energy $E_{C\Sigma}=(2e)^2/(C_J+C_L+C_R)$, where $C_{L,R}$ are the capacitances on the left and right branch.

When the Josephson energies are smaller than the inductive energies one can neglect the last term in $H_l$, and the eigenstates of the normal modes are harmonic oscillator number states $\ket{l_1,l_2}$.
In this perturbative limit, $\beta_L\lesssim 1$ and $\alpha,\beta \lesssim 1/\beta_L$, we can trace out the oscillator degrees of freedom.
Integrating $H_l$ over the harmonic oscillator wave functions, the effective BiSQUID Hamiltonian in the presence of inductance is
\begin{align*}
  H_l \approx \hbar\omega_1&(l_1+1/2)+\hbar\omega_2(l_2+1/2) + E_{C\Sigma}(n-n_g)^2+ \\
  -E_J  [\, &\rho_0\cos \theta + \alpha \rho_L\cos(\theta - \varphi_L) + \beta \rho_R\cos(\theta + \varphi_R) ].
\end{align*}

The Josephson energies of the junctions are renormalized by exponential factors
\begin{align*}
  \rho_0 &= e^{-\gamma_1^2(2l_1+1)z_1/4-\gamma_2^2(2l_2+1)z_2/4} \\
  \rho_L &= e^{-\gamma_3^2(2l_1+1)z_1/4-\gamma_4^2(2l_2+1)z_2/4} \\
  \rho_R &= e^{-\gamma_5^2(2l_1+1)z_1/4-\gamma_6^2(2l_2+1)z_2/4}.
\end{align*}
They depend not only on the participation ratios and impedances of the harmonic modes, but also on the photon occupation, resulting in a smaller topologically nontrivial phase as $l_1,l_2$ are increased.

\begin{figure}[ht]
    \includegraphics[width=\columnwidth]{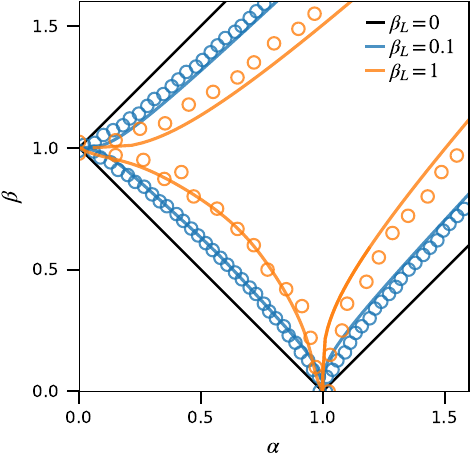}%
    \caption{\label{fig:PhaseDiagramBetaL} 
    BiSQUID phase diagram in the presence of inductance $E_J/\beta_L$ on each branch.
    The critical line separating the topologically trivial and nontrivial phases is shown for $\beta_L=0$ (zero inductance, black), $\beta_L=0.1$ (blue), and $\beta_L=1$ (orange).
    The solid lines for $\beta_L>0$ are determined from the perturbative model (\cref{app:betal}).
    The circle markers are obtained by numerical diagonalization of the BiSQUID Hamiltonian including inductance.
    The plasma frequency of each junction is kept constant for all $\alpha$ and $\beta$.
    The capacitance of the middle junction is fixed to $C_J=(8e^2L_J)/(\varphi_0^2)$, with $L_J$ the Josephson inductance of the middle junction and $\varphi_0=\hbar/2e$ the reduced flux quantum.
    }
\end{figure}

We can now solve for the Josephson degrees of freedom $n,\theta$, with the only difference being the renormalization of the Josephson energies.
The triangle inequality $|\alpha \rho_L-\beta \rho_R|<\rho_0<\alpha \rho_L+\beta \rho_R$ can still be satisfied, but the position of the critical lines in the phase diagram will be shifted.
Since $\varphi_{L,R}$ is either $0$ or $\pi$ on the critical lines, one can solve the transcendental equations $\rho_0\pm\alpha \rho_L\pm\beta \rho_R=0$ to determine their positions.
The solutions of these equations for the ground states $l_1=l_2=0$ and $\beta_L= 0.1, 1$ (solid lines) show excellent agreement with the numerically obtained critical lines in the region $\alpha,\beta \lesssim 1/\beta_L$~[\cref{fig:PhaseDiagramBetaL}].
We use the python package scQubits~\cite{groszkowskiScqubits2021,chittaComputeraided2022} to numerically diagonalize the BiSQUID Hamiltonian with inductance, determining the boundary between topologically trivial and nontrivial regions (circle markers) by computing the energy gap $f_{01}$ for $n_g=1/2$.

\section{\label{app:fab}Fabrication}

The devices are fabricated on \SI{350}{\um}-thick wafers of high-resistivity intrinsic silicon capped with a \SI{150}{\nm} Si02 surface oxide. 
We begin by cleaning the substrate using a soft oxygen plasma etching step in a reactive ion etcher.
Next, we deposit a layer of aluminum on the entire wafer. 
This aluminum is deposited at a rate of 1 nm/s using a high-vacuum Plassys electron gun evaporator.
Before exposing the wafer to ambient atmosphere, the aluminum is passivated in the evaporator in 200 mbar of pure dioxygen for 10 minutes.

In a first optical lithography step, we define the resonator, control lines, and alignment structures. 
To make the negative mask, we spin coat a layer of S1813 resist, bake it at 115°C for one minute, expose it using a LW405B laser writer with a dose of about 200 mJ/cm$^2$, and develop it in MF319 developer for one minute.
To remove the aluminum in exposed areas, we perform a wet etch, which involves first hard baking the resist at 150°C for 3 minutes, followed by immersing the wafer in MF319 for 5 to 7 minutes and rinsing it with water.
Then, we dice the wafer in individual \qtyproduct[product-units=power]{10 x 10}{\mm} dies.

Next, we fabricate BiSQUID's Josephson junctions and test structures using electron-beam lithography.
We spin coat a bilayer of MMA EL11 and PMMA 950 A4 at 3000 rpm for 60 seconds each.
The lithography mask is patterned using a 30 kV FEI Magellan scanning electron microscope equipped with Raith ELPHY quantum software.
The typical Josephson junction area is  \qtyproduct[product-units=power]{200x200}{nm} with a critical current density of \SIrange{150}{200}{\nA \per \um \squared}.
The mask is developed in a 1:3 MIBK/IPA mixture for 55 seconds and rinsed in IPA for 60 seconds.
The Josephson junctions are deposited using a Plassys electron-beam evaporator and a double-angle Dolan evaporation technique.
Residual resist is removed by a soft oxygen plasma etch.
Then, a first layer of 12 nm of aluminum is deposited at a rate of 0.1 nm/s and an angle of 30°.
The tunnel barrier is formed by exposing the junctions to 200 mbar of pure dioxygen for 10 minutes.
A second layer of 80 nm of aluminum is deposited at a rate of 1 nm/s and an angle of $-30$°. 
The difference in thickness between the first and second electrodes leads to a superconducting gap difference across the junctions, which can prevent quasiparticles from tunneling across the junction (\cref{app:QP}).
The junctions are finally passivated in 200 mbar of pure dioxygen for 10 minutes.
The resist is stripped using NMP 1165 for a few hours at 80°C.


The BiSQUID leads are capacitively coupled to the ground plane over an area of approximately $\SI{800}{\micro\meter\squared}$.
We do not remove the aluminum oxide of the ground plane, so the connection is capacitive with a high capacitance.
However, for some devices (SQUID \emph{R16} and BiSQUID R17C), an extra step has been added to connect test JJs to measurement pads.
We use the same resist as for the first lithography and open small patching areas around test JJs leads and measurement pads. 
The resist is then hard-baked at 150°C for 3 minutes.
In a Plassys electron-beam evaporator, we mill the aluminum oxide in two 4-minute argon milling steps with an acceleration voltage of 500 V and a current of 30 mA.
We then deposit 100 nm of aluminum at a rate of 1 nm/s and passivate the aluminum by 10 min oxidation in 200 mbar of pure dioxygen.

\section{\label{app:expsetup}Experimental setup}

\begin{figure*}
\includegraphics[width=.8\textwidth]{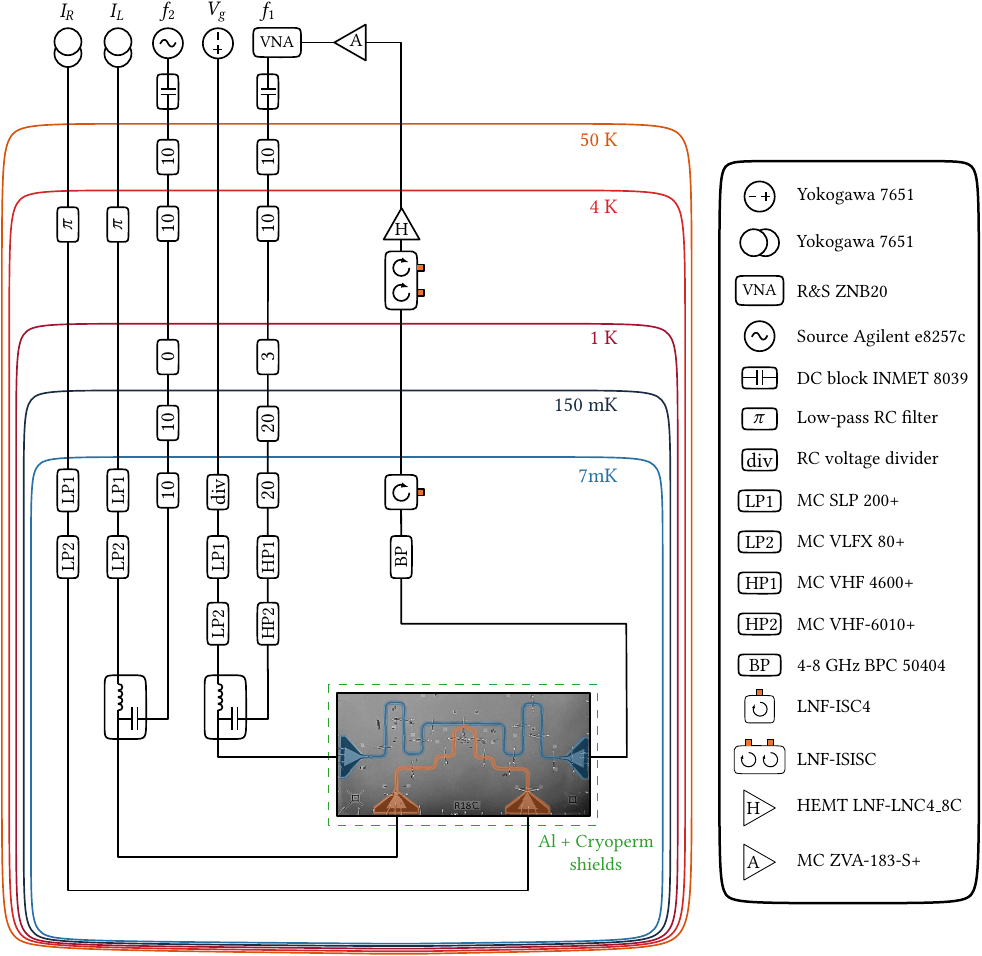}%
    \caption{\label{fig:full_wiring}
    Full wiring of the cryostat for two-tone spectroscopy experiments.
    NbTi twisted pair wires are used for dc currents $I_L$ and $I_R$.
    An in-house flexible polyimide cable is used for the gate voltage $V_g$.
    It supports both distributed and lumped low-pass RC filtering at all stages of the fridge.
    }
\end{figure*}

The simplified wiring used to perform two-tone spectroscopy is shown in~\cref{fig:full_wiring}.
A VNA is used to generate a signal at a frequency $f_1$ to probe the resonator.
This signal is combined at the mixing chamber in a bias tee with a dc gate voltage $V_g$ generated by a Yokogawa 7651 voltage source.
The dc gate voltage is filtered by custom-made flexible cables with both lumped and distributed filtering, resulting in a low-pass cutoff of around 50 Hz and a voltage attenuation factor of 17.
A second tone at a frequency $f_2$ is produced by an Agilent signal generator and combined with the current $I_X$ in a second bias tee which is connected to the flux line.
We use NbTi twisted pairs for the dc currents, with attenuation at the 4K stage and additional low-pass filtering at the mixing chamber stage.
Microwave input signals are attenuated at different stages of the cryostat, with a total of around \qty{-75}{dB} for the $f_1$ line and \qty{-55}{dB} for the $f_2$ line. 
Low-pass and high-pass microwave filters are used on the dc and microwave lines, respectively.
Later in the experiment, we also added Eccosorb filters on all dc and rf lines at the mixing chamber stage.
However, we observe no significant change, and quasiparticle poisoning rates are unaffected by this additional filtering.

The output signal is amplified first at 4K with a LNF cryo-HEMT with a gain of 42 dB and again at room temperature with a Mini-Circuits ZVA amplifier with a gain of 26 dB.
A Josephson traveling-wave parametric amplifier is also used for some measurements~\cite{macklinQuantumlimitedJosephsonTravelingwave2015}. 
The BiSQUID device is isolated from amplifier noise by a Microtronics band-pass filter, a single isolator at the mixing chamber and a double circulator at 4K.
The output signal is recombined with the reference signal in the VNA, enabling the transmission parameter $S_{21}$ to be measured.


The measured resonator and BiSQUID parameters are given in~\cref{tab:all_param}.
To characterize the resonators, we fit the transmission data to extract the loaded quality factor $Q_L$.
We use the FastCap tool in LayoutEditor to estimate the external coupling capacitances from the CPW resonator to the feedlines.
From these capacitances, we estimate the external quality factor $Q_\mathrm{ext}$.
The output capacitance is designed to be larger than the input capacitance such that most of the photons from the resonator are emitted into the output feedline.
The CPW resonator is designed to have an impedance of 75 Ohm to increase its coupling with BiSQUIDs.
The coupling of a BiSQUID and CPW resonator is estimated from vacuum Rabi splitting measurements.
We find typical values of $g=2\pi\,20$ MHz.
Further details are given in Ref. \cite{leo_thesis}.


\section{\label{app:QP}Quasiparticle poisoning}

\begin{figure}[ht]
 \includegraphics[width=\columnwidth]{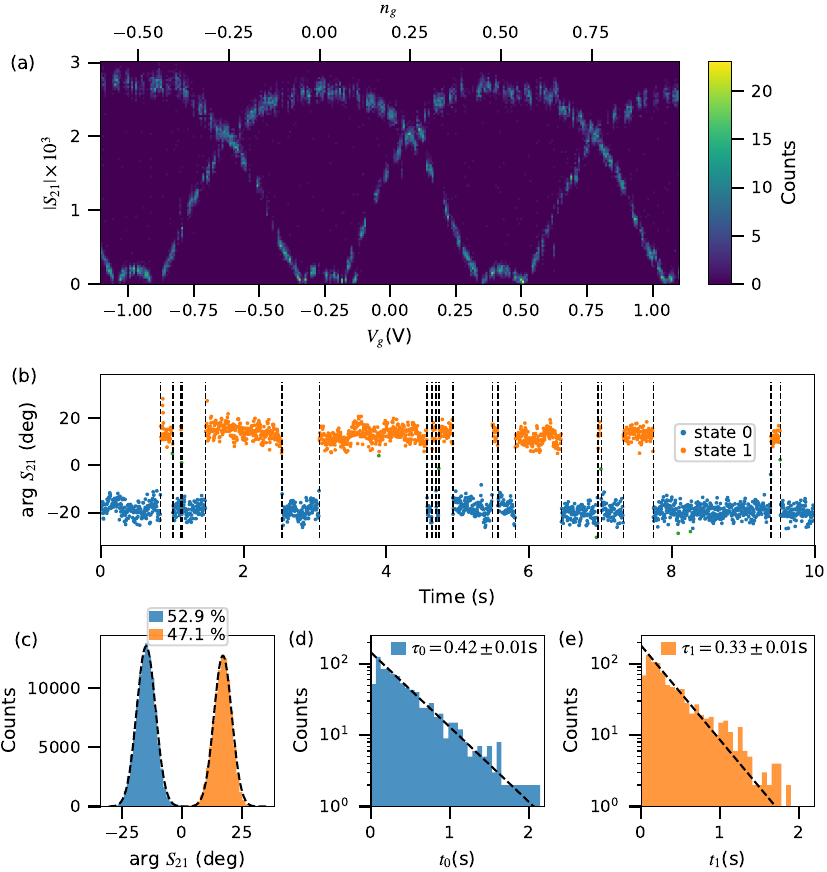} 
    \caption{\label{fig:QP_poisoning} 
    Quasiparticle poisoning in BiSQUID R18.
    (a) Measured magnitude of the resonator transmission at fixed frequency as a function of gate voltage $V_g$ for fixed $\varphi_L=\varphi_R=0$.
    For each value of $V_g$, we collect 50 data points with an individual averaging time of $\SI{5}{\milli\second}$, and we indicate counts using the color scale.
    The two roughly sinusoidal lines shifted by approximately $\SI{0.68}{\volt}$ correspond to the two QP parities.
    Values of $V_g$ are given after attenuation in the measurement lines.
    (b) First 10 seconds of an 800-second measurement of the resonator with an averaging time of $\SI{5}{\milli\second}$ per point. 
    The reduced flux threading the SQUID is set to $\varphi_\mathrm{diag}\approx\pi$, and the charge offset is set to $n_g \approx 0.5$.
    Data points are separated into two parity states (blue and orange); then, parity switching events are identified (black dashed lines).
    Outlier data (green markers) are not included in the analysis.
    (c) Histogram of the phase of the full time trace fitted to two Gaussian peaks.
    We find two states with occupation probabilities of 53$\%$ for state 0 (blue) and 47$\%$ for state 1 (orange).
    (d), (e) Histograms of occupation times $t_0$ of state 0 (d) and $t_1$ of state 1 (e), fitted to exponential distributions with time constants $\tau_0$ and $\tau_1$.
    The vertical axes indicate counts in logarithmic scale.
    }
\end{figure}

The presence of quasiparticles (QPs), or unpaired electrons, is harmful to many superconducting devices \cite{glazmanBogoliubov2021, wilenCorrelated2021}.
The origin of these nonequilibrium QPs is still unclear, but possible explanations include stray infrared photons and ionizing radiation from environmental radioactive sources and cosmic rays \cite{barendsMinimizing2011,vepsalainenImpact2020,cardaniReducing2021,gordonEnvironmentalRadiationImpact2022}.

In the absence of QPs, the energy of a Josephson junction is periodic with the charge offset $2 e n_g$.
However the tunneling of a QP through the junction changes the parity of the superconducting island and shifts the charge offset by $1e$.
If QPs are tunneling to the island faster than they recombine or tunnel out, the parity of the island will be constantly changing.
By measuring faster than the parity switching time, we can observe the typical ``eye-pattern'' consisting of two states with different quasiparticle parity and $1e$ charge offset periodicity [\cref{fig:QP_poisoning}(a)].

To extract the typical QP parity switching time, we measure the state of BiSQUID device R18 at $\varphi_\mathrm{diag}\approx\pi$ and $n_g=0.5$ for 800 seconds.
The first 10 seconds of the time trace are shown in \cref{fig:QP_poisoning}(b).
We identify a parity switching event if the state of the system changes for at least three data points.
From the histogram of the full time trace, we can clearly identify two states with similar average occupation of 50$\%$ [\cref{fig:QP_poisoning}(c)].
We then extract all the occupation times $t_i$ of state $i$ and fit their histogram to an exponential distribution $f(t_i)=A\exp(-t_i/\tau_i)$ which yields time constants of about $\SI{0.4}{\second}$. 
We measure QP switching times at other values of $\varphi_L,\varphi_R$ and $n_g$ and find consistent switching times between 0.1 and 1 second.

\begin{figure}[ht!]
    \includegraphics[width=.49\textwidth]{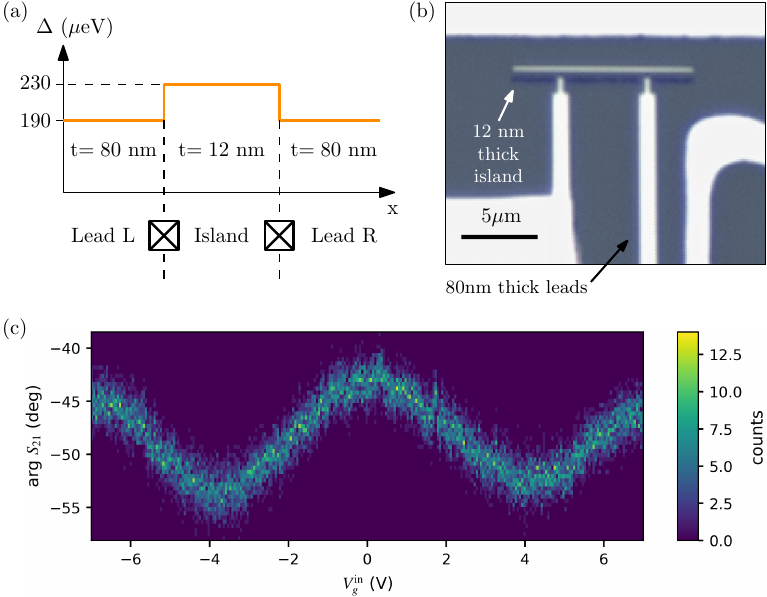}%
    \caption{\label{fig:QP_gap_engineering}
    Superconducting gap engineering using varying aluminum thicknesses.
    (a) The $\SI{12}{\nano\meter}$ thick island has a larger superconducting gap than the rest of the circuit.
    The extra energy of $\delta\Delta = \SI{40}{\micro\eV}$ can partially prevent QP from entering the island.
    (b) Optical microscope image of a SQUID with a $\SI{12}{\nano\meter}$ thick island (faint dark blue) and $\SI{80}{\nano\meter}$ thick leads (white).
    The thin island can only be seen with a large brightness.
    The mirror image of the island (white) is due to the angle evaporation technique, but this island is not connected to the rest of the circuit.
    (c) Resonator transmission measured as in \cref{fig:QP_poisoning} (a) but with the gap-engineered SQUID \emph{R16}.
    Only a single quasiparticle state is observed, meaning that QP poisoning events are strongly suppressed.
    However, for several other fabrications of gap-engineered SQUIDs and BiSQUIDs with similar recipes and designs, the gap-engineering technique did not work and we observed  typical QP parity switching times $\tau_i$ of \SIrange{0.1}{0.5}{s}.
    }
\end{figure}

We increase the superconducting gap $\Delta$ of the BiSQUID island by reducing the thickness of the aluminum to \SI{12}{\nm}, leading to an increase of about \SI{40}{\micro\eV} of $\Delta$ compared to the \SI{80}{\nm} thick aluminum layer used for the rest of the BiSQUID (\cref{fig:QP_gap_engineering}).
This process creates an energy barrier, potentially preventing QP from tunneling to the island \cite{aumentadoNonequilibrium2004,yamamotoParity2006,courtEnergy2008,riwarEfficient2019, kalashnikovBifluxon2020, catelaniUsing2022}.
Gap-engineering works well in some devices, with a typical QP tunneling time of several seconds and an average population of poisoned parity of only a few percent.
However, the effectiveness of gap engineering seems to be highly fabrication dependent and, in most cases, does not function properly.
One possible explanation for the failure of gap engineering is the presence of defects in the oxide barrier of the junctions, which may form spatially localized subgap states \cite{kurterQuasiparticleTunnelingProbe2022}.
We tried adding Eccosorb filters to all dc and rf lines but did not observe any impact on QP poisoning.
Increasing the attenuation of Eccosorb filtering and improving the sample shielding with tight indium seals may help suppress the high-frequency photon flux and reduce quasiparticle poisoning \cite{connollyCoexistenceNonequilibriumDensity2024}.

In addition to quasiparticle parity switching, BiSQUIDs are also sensitive to charge noise, which may be attributed to fluctuating two-level systems in dielectrics or surface defects~\cite{wangSurfaceParticipationDielectric2015}.
By monitoring the charge offset over time, we observe random amplitude jumps approximately every 10 minutes.
We also observe a slow drift of the charge offset on the order of $0.1\%$ of a $2e$ period per minute \cite{leo_thesis}.
To maintain a constant value of charge offset over a long measurement, we regularly perform quick calibration measurements.
We probe the resonator while sweeping the charge offset of the BiSQUID over 1.5 periods. 
We set the reduced fluxes $\varphi_L,\varphi_R$ of the BiSQUID such that its energy is close to the resonator without crossing it.
This method allows an easier fitting procedure to extract the correspondence between $V_g$ and $n_g$.
This calibration measurement takes about 5 seconds, and we perform it every \SIrange{1}{2}{\min} to keep track of the charge offset.
If we detect a small drift during a measurement, we adjust the charge offset accordingly, and if we detect a jump, we discard the last data and remeasure.

\begin{figure}[t]
   \includegraphics[width=\columnwidth]{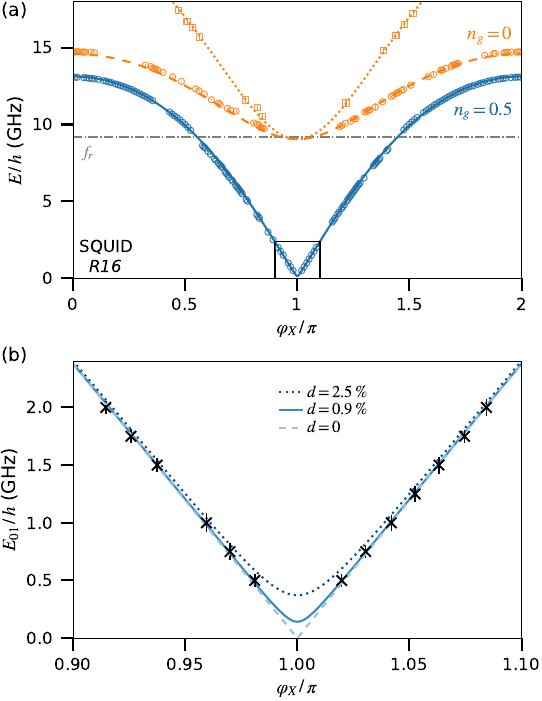}%
    \caption{\label{fig:squid_data}
    SQUID \emph{R16} energy transitions as a function of reduced flux $\varphi_X$ extracted from two-tone spectroscopy measurements.
    (a) Transition $f_{01}$ measured at charge offset $n_g=0.5$ (blue circles along solid lines) and $n_g=0$ (orange circles along dashed lines).
    The transition $f_{02}$ is measured at $n_g=0$ (orange squares along dotted line) and is out of the measurement range at $n_g=0.5$.
    Error bars indicate the measured FWHM of transitions, and lines are best fits to the data using the BiSQUID Hamiltonian with $\beta=0$.
    The resonator bare frequency $f_r=9.178$ GHz is indicated by a gray dash-dot line.
    (b) Zoom of the $f_{01}$ transition at $n_g=0.5$ around $\varphi_X=\pi$. 
    The best fit of the data yields the asymmetry parameter $d=(\alpha-1)/(\alpha+1)=\qty{0.9}{\percent}$ (solid line).
    The other two lines show simulations for an upper bound of the asymmetry of $d=\qty{2.5}{\percent}$ (dotted line) and a fully symmetric SQUID with $d=0$ (dashed line).
    }
\end{figure}

\section{\label{app:sqdata}SQUID data}

We also performed the spectroscopy of a SQUID as a control experiment.
Two external parameters control SQUID energy levels: one reduced flux $\varphi_X$ and one charge offset $n_g$.
At half-flux quantum $\varphi_X=\pi$, a perfectly symmetric SQUID reduces to the charging Hamiltonian $H_C$, which has a hidden quantum mechanical supersymmetry (\cref{fig:intro}).
However, any asymmetry between left and right junctions prevents the cancellation of the effective Josephson energy $E_j^*$, which lifts all the degeneracies of the energy levels.

The SQUID circuit \emph{R16} was designed to be symmetric with equal Josephson energies; however, the fabrication process introduces a small asymmetry between left and right junctions. 
The spectroscopy of SQUID \emph{R16} is shown in~\cref{fig:squid_data} as a function of $\varphi_X$ at two values of $n_g$.
At $n_g=0$, we measure the first two transitions $f_{01}$ (orange circles) and $f_{02}$ (orange squares).
They converge to a similar value at $\varphi_X=\pi$.
However, we cannot resolve the transitions near this point because it is almost resonant with the readout resonator frequency $f_r$ (gray dash-dot line).
The first transition $f_{01}$ is also measured at $n_g=0.5$, and a zoom down to $\SI{0.5}{\GHz}$ is shown in~\cref{fig:squid_data}(b).
Similarly to the BiSQUIDs, the spectral line of $f_{01}$ disappears near $\varphi_X=\pi$ because of thermal effects and because of the decrease of the charge-coupled resonator shift.

We fit the transitions to the eigenenergies of the BiSQUID Hamiltonian \cref{eq:full_H} with $\beta=0$.
We use three free parameters: the Josephson energy $E_j$, the charging energy $E_C$, and the asymmetry parameter $d=(\alpha-1)/(\alpha+1)$.
The best-fit parameters are given in~\cref{tab:all_param}, and the simulated energy levels are shown in~\cref{fig:squid_data}.
We find an asymmetry parameter of $d=\qty{0.9}{\percent}$.
Additional simulated energy levels are shown in~\cref{fig:squid_data}(b) for an upper bound of the asymmetry of $d=\qty{2.5}{\percent}$ and for a fully symmetric SQUID with $d=0$.
Measuring $f_{01}$ down to lower frequencies would be required to further constrain the asymmetry parameter.


%




\begin{acknowledgments}
  We thank Joël Griesmar and Jean-Damien Pillet for help conceiving this project and Jean-Noël Fuchs for elucidating the analogy to topological semimetals.
  We thank Jeffrey Teo, Roman Riwar, Valla Fatemi, Hugues Pothier, Landry Bretheau, Benoît Douçot, and Benjamin Wieder for discussions.
  This project has received funding from the European Research Council (ERC) under the European Union's Horizon 2020 research and innovation programme (Grant Agreement No. 636744).
  The research was also supported by Japanese Society for the Promotion of Science International Research Fellowship L22559, IDEX Grant No. ANR-10-IDEX-0001-02 PSL, a Paris ``Programme Emergence(s)'' Grant, the Office of Naval Research under Award No. N00014-20-1-2356, and the Thomas Jefferson Fund, a program of FACE Foundation launched in collaboration with the French Embassy.
  We thank IARPA and MIT Lincoln Labs for providing a TWPA and ESPCI Paris for providing SEM services.
%
\end{acknowledgments}
\clearpage

\bibliography{main.bib}

\end{document}